\documentclass[12pt,a4paper]{article}

\usepackage{a4wide}
\usepackage{amsmath}
\usepackage{bm}
\usepackage{amssymb}
\usepackage{hyperref}
\usepackage{epic}
\usepackage{mathrsfs}
\usepackage{color}

\newcommand{\p}{\partial}
\newcommand{\Tr}{{\rm Tr \,}}
\newcommand{\Op}{\mathcal{O}}
\newcommand{\fldZ}{\mathcal{Z}}

\newcommand{\fldD}{\mathcal{D}}

\newcommand{\alg}[1]{\mathfrak{#1}}
\newcommand{\superN}{\mathcal{N}}
\newcommand{\Baxter}{P^{(0)}_M}
\newcommand{\DBaxter}{P^{(0)'}_M}
\newcommand{\Baxtersub}{P^{(2)}_M}
\newcommand{\DBaxtersub}{P^{(2)'}_M}
\newcommand{\ex}{\Upsilon}
\newcommand{\sca}{S_{0}}
\newcommand{\mat}{S_{\boxplus}}
\newcommand{\intd}{Y_{Q}}
\newcommand{\dress}{S_{\sigma}}

\def\HBS{\mathbb S}
\def\HS{S}
\def\reciP{\mathcal{P}}
\def\dotgamma{\dot \gamma}
\def\ddotgamma{\ddot \gamma}
\def\dddotgamma{\dddot \gamma}
\def\ddddotgamma{\dddot \gamma}

\newcommand{\gym}{g\indups{YM}}
\newcommand{\indups}[1]{_{\mathrm{\scriptscriptstyle #1}}}
\newcommand{\gaba}{\gamma\inddowns{ABA}}
\newcommand{\inddowns}[1]{^{\mathrm{\scriptscriptstyle #1}}}

\newcommand{\beq}{\begin{equation}}
\newcommand{\eeq}{\end{equation}}
\newcommand{\beqa}{\begin{eqnarray}}
\newcommand{\eeqa}{\end{eqnarray}}
\newcommand{\nn}{\nonumber}

\newcommand{\sfrac}[2]{{\textstyle\frac{#1}{#2}}}

\newcommand{\M}{M}

\makeatletter
\def\mr@ignsp#1 {\ifx\:#1\@empty\else #1\expandafter\mr@ignsp\fi}%
\newcommand{\multiref}[1]{\begingroup
\xdef\mr@no@sparg{\expandafter\mr@ignsp#1 \: }%
\def\mr@comma{}%
\@for\mr@refs:=\mr@no@sparg\do{\mr@comma\def\mr@comma{,}\ref{\mr@refs}}%
\endgroup}
\makeatother

\numberwithin{equation}{section}

\begin{document}
\thispagestyle{empty}

\begin{flushright}\footnotesize

\texttt{Imperial-TP-AR-2009-4}\\
\vspace{0.5cm}
\end{flushright}
\setcounter{footnote}{0}

\begin{center}
{\Large{\bf Five-Loop Anomalous Dimension of Twist-Two Operators}}
\vspace{15mm}

{\sc T.~\L ukowski $^a$, A.~Rej $^b$, V.~N.~Velizhanin $^c$}\\[5mm]

{\it $^a$  Institute of Physics, Jagellonian University,\\
 ul. Reymonta 4, 30-059 Krak{\'o}w, Poland}\\[5mm]

{\it $^b$ Blackett Laboratory, Imperial College, London SW7 2AZ, U.K.}\\[5mm]

{\it $^c$ Theoretical Physics Department\\
Petersburg Nuclear Physics Institute\\
Orlova Roscha, Gatchina\\
188300 St.~Petersburg, Russia}\\[5mm]

\textbf{Abstract}\\[2mm]
\end{center}

\noindent{In this article we calculate the five-loop anomalous dimension of twist-two operators in the planar $\superN=4$ SYM theory. Firstly, using reciprocity, we derive the contribution of the asymptotic Bethe ansatz. Subsequently, we employ the first finite-size correction for the $AdS_5\times S^5$ sigma model to determine the wrapping correction. The anomalous dimension found in this way passes all known tests provided by the NLO BFKL equation and double-logarithmic constraints. This result thus furnishes an \textit{infinite} number of experimental data for testing the veracity of the recently proposed spectral equations for planar AdS/CFT correspondence.}
\newpage

\setcounter{page}{1}
\def\zt{\zeta(3)}
\def\zf{\zeta(5)}
\def\zs{\zeta(7)}

\section{Introduction and Summary}
\label{sec:intro}

Integrable structures appeared for
the first time in four-dimensional gauge field theories
in the context of high-energy scattering in quantum chromodynamics (QCD).
In the Regge limit the
scattering amplitudes of colorless particles are dominated by the exchange of two effective particles,
termed reggeized gluons. A compound of two of these
particles is frequently called the {\it pomeron} and corresponds to the leading asymptotics of the scattering amplitudes in the Regge limit. In the infinite energy limit $s\to \infty$, however, the Froissart bound on the cross-section applies and further corrections must be taken into account in order to comply with unitarity. Physically this corresponds to including part of the interactions between \textit{many} reggeized gluons. In the planar limit the Hamiltonian governing the dynamics of these gluons turns out to be integrable \cite{Lipatov:1993yb}. Moreover, it can be decomposed into a holomorphic and an antiholomorphic part, each of which can be identified with the Heisenberg magnet of spin $0$, see \cite{Lipatov:1994xy,Faddeev:1994zg}.  The length of this
spin chain equals the number of reggeized gluons considered.
The pomeron corresponds to the ground state of the length-two spin chain.
Although formally the S-matrix of the $\superN=4$ supersymmetric Yang-Mills (SYM) theory cannot be defined, the analysis of the scattering amplitudes reveals that gluons also reggeize in this case\footnote{This feature is common to many gauge theories with non-abelian gauge group.}. On top of that,  the large amount of symmetries significantly simplifies the reggeon Hamiltonian and the Balitsky-Fadin-Kuraev-Lipatov (BFKL)~\cite{Lipatov:1976zz,Kuraev:1977fs,Balitsky:1978ic} equation, i.e. the eigenvalue equation for the reggeon Hamiltonian, takes a particularly simple form \cite{Kotikov:2002ab}.

The planar $\superN=4$ gauge theory exhibits a very remarkable property, absent in most gauge theories\footnote{See, however, \cite{Braun:1998id} and \cite{Belitsky:2005bu} for the one- and two-loop
integrability in $\mathfrak{sl}(2)$ sector of QCD.}. Namely, the dilatation operator is believed to be asymptotically integrable. In the groundbreaking paper \cite{Minahan:2002ve} the one-loop integrability of the dilatation operator in certain subsectors of the gauge side of the correspondence was discovered. Later on \cite{Beisert:2003jj}, the complete one-loop dilatation operator and the one-loop Bethe equations were written down in \cite{Beisert:2003yb}. After many non-trivial steps \cite{fussnote} the form of the all-loop asymptotic Bethe equations (ABE) was conjectured \cite{Beisert:2005fw} up to the so called dressing factor, which only contributes starting from the four-loop order. Subsequently, relying on the crossing equation proposed in \cite{Janik:2006dc} and assuming certain transcendentality properties, it was possible to uniquely fix this factor \cite{Beisert:2006ez}. This completed the solution to the spectral problem of the planar $\mathcal{N}=4$ SYM theory in the asymptotic region. The asymptoticity of these equations means that for a generic operator with $L$ constituent fields the corresponding anomalous dimension can be calculated correctly up to the $\mathcal{O}(g^{2L})$ order.

Beyond this order the asymptotic Bethe equations are \textit{not} valid, as it was shown in \cite{Kotikov:2007cy}. In \cite{Kotikov:2007cy} the anomalous dimension of twist-two operators has been calculated in a closed form at fourth order in perturbation theory. The analytic continuation to  negative integer values of the spin allowed for a comparison of the results with the predictions of the BFKL equation. As a result, a maximal violation of the BFKL prediction has been found, which unambiguously corroborates the breakdown of the asymptotic integrability. The latter was established by disregarding the contribution of a certain class of Feynman diagrams while studying the structure of mixing of the operators. These diagrams are commonly referred to as wrapping diagrams, the reason being their topological properties \cite{Sieg:2005kd}. In the spin chain picture these diagrams correspond to the interactions wrapping around the spin chain. Since the interaction between two nearest neighbours provides a factor of $g^2$, the first wrapping diagram may appear at the order $\Op(g^{2 L})$, where $L$ is the length of the spin chain. For twist-two (length-two) operators the supersymmetry \textit{delays} the wrapping interactions to the four-loop order. In that sense the breakdown of the ABE was expected from this point of view.

In the seminal paper \cite{Bajnok:2008bm} it was proposed to evaluate the L\"uscher corrections, previously proposed for the $AdS_5 \times S^5$ sigma model in \cite{Janik:2007wt}, to calculate the leading wrapping corrections to the simplest unprotected operator in the $\mathfrak{sl}(2)$ sector, the Konishi operator. The result remarkably coincides with very complicated field theory computations \cite{Fiamberti:2007rj,Velizhanin:2008jd}. In the same way the four-loop wrapping correction for the whole family of twist-two operators has been determined \cite{Bajnok:2008qj}. When added to the ABA result, it restores the agreement with the BFKL prediction! Recently, the same correction has been employed to calculate the five-loop anomalous dimension of the Konishi operator \cite{Bajnok:2009vm}.

As for other integrable sigma models, the full solution to the spectral problem is believed to be given in terms of a set of TBA equations. There has been recently a lot of progress in this direction \cite{Ambjorn:2005wa}-\cite{Arutyunov:2009zu} and ultimately the Y-system \cite{Gromov:2009tv} and the TBA equations for the ground state  \cite{Bombardelli:2009ns}-\cite{Arutyunov:2009ur} have been formulated. The authors of \cite{Gromov:2009bc} have also proposed the TBA equations for excited states in the $\mathfrak{sl}(2)$ sector.

It is thus necessary to check these all-loop proposals at higher orders of perturbation theory\footnote{In \cite{Gromov:2009zb} TBA equations for the Konishi operator have been studied numerically at many different values of the coupling constant. In particular, the strong-coupling expansion of the scaling dimension has been conjectured, which structurally but not quantitatively coincides with the string theory computation \cite{Roiban:2009aa}. As shown in \cite{Rej:2009dk}, the contribution of the non-wrapping diagrams violates the structure of the strong-coupling expansion and thus analytic derivation of the strong-coupling expansion seems necessary. This is even more urging in view of the fact that the TBA equations seem to predict infinitely many singular values of the coupling constant, see \cite{Arutyunov:2009ax}. }. In this article we derive the \textit{full} five-loop anomalous dimension of twist-two operators, which may serve as the sought-for ``experimental data'' to verify the proposed spectral equations of $\superN=4$ SYM. It can be split into the ABA part and the contribution of the wrapping interactions
\beq \label{totalad}
\gamma_{10}(M)=\gamma^{\textrm{ABA}}_{10}(M)+\Delta_w(M)\,.
\eeq
We present the explicit expression for $\gamma^{\textrm{ABA}}_{10}(M)$ in appendix \ref{sec:ABAresult}. For the wrapping correction $\Delta_w(M)$ we have found
\beqa
\Delta_w&=&
13440\, \zs \HS_1^2
-1536\, \zt^2 \HS_1^3
+2560\, \zf \HS_1\Big(
3\,\HS_1(2\,\HS_{-2}+\HS_2)
-\HS_1^3
+\HS_{-3}
+\HS_3
-2\,\HS_{-2,1}\Big)\nn\\
&&+1024\, \zt \HS_1
\Big(-2\,  \HS_1^3 \HS_{-2}
   + 2\,\HS_1^2 (2\, \HS_{-3}+3\, \HS_3)
   +\HS_1(4\, \HS_{-2}^2
   +6\, \HS_2 \HS_{-2}
   +3\, \HS_{-4}
   -\HS_4\nn\\
&&   -2\, (\HS_{-3,1}-2\, \HS_{-2,-2}+\HS_{-2,2}+\HS_{3,1}-2\, \HS_{-2,1,1})
    )
  +2\, \HS_{-2} (\HS_{-3}+\HS_3-2\, \HS_{-2,1})
\Big)\nn\\
&&-1024\, \HS_1 \Big(
  (\HS_1 (3\, \HS_2 +2\, \HS_{-2}) + \HS_{-3}+\HS_3-2\, \HS_{-2,1}-\HS_1^3)
     (\HS_{-5}-\HS_5+2\,\HS_{-2,-3}-2\,\HS_{3,-2}\nn\\
&&+2\,\HS_{4,1}-4\, \HS_{-2,-2,1})
 +2\, \HS_1^2( 2 \HS_{-6}
  -2\, \HS_6
  -\HS_{-4,-2}
  +2\, \HS_{-3,-3}
  +3\, \HS_{-2,-4}
  +\HS_{-2,4} \nn\\
&&  -2\, \HS_{3,-3}
  -2\, \HS_{4,-2}
  +\HS_{4,2}
  +4\, \HS_{5,1}
  -4\, \HS_{-3,-2,1}
  -4\, \HS_{-2,-3,1}
  -2\, \HS_{-2,-2,-2}
  -2\, \HS_{-2,-2,2})\nn\\
&& +  \HS_1(5\, \HS_{-7}
  -5\, \HS_7
  -4\, \HS_{-6,1}
  +4\, \HS_{-5,-2}
  -\HS_{-5,2}
  +3\, \HS_{-4,-3}
  +\HS_{-3,-4}
  -\HS_{-3,4}
  +8\, \HS_{-2,-5}\nn\\
&&  -6\, \HS_{-2,5}
  -4\, \HS_{3,-4}
  +2\, \HS_{3,4}
  -8\, \HS_{4,-3}
  +3\, \HS_{4,3}
  -6\, \HS_{5,-2}
  +\HS_{5,2}
  +6\, \HS_{6,1}
  +2\, \HS_{-5,1,1}\nn\\
&&  -6\, \HS_{-4,-2,1}
  -2\, \HS_{-3,-3,1}
  +2\, \HS_{-3,-2,-2}
  -2\, \HS_{-3,1,-3}
  -8\, \HS_{-2,-4,1}
  +6\, \HS_{-2,-3,-2}
  -2\, \HS_{-2,-3,2}\nn\\
&&  +14\, \HS_{-2,-2,-3}
  -6\, \HS_{-2,-2,3}
  -2\, \HS_{-2,1,-4}
  +2\, \HS_{-2,1,4}
  -2\, \HS_{-2,2,-3}
  -4\, \HS_{-2,3,-2}
  +10\, \HS_{-2,4,1}\nn\\
&&  +2\, \HS_{3,-3,1}
  -4\, \HS_{3,-2,-2}
  +2\, \HS_{3,-2,2}
  +2\, \HS_{3,1,-3}
  +2\, \HS_{3,2,-2}
  +10\, \HS_{4,-2,1}
  +6\, \HS_{4,1,-2}
  -2\, \HS_{4,1,2}\nn\\
&&  -2\, \HS_{4,2,1}
  -2\, \HS_{5,1,1}
  +4\, \HS_{-3,1,-2,1}
  +4\, \HS_{-2,-3,1,1}
  -20\, \HS_{-2,-2,-2,1}
  -8\, \HS_{-2,-2,1,-2}\nn\\
&&  +4\, \HS_{-2,-2,1,2}
  +4\, \HS_{-2,-2,2,1}
  +4\, \HS_{-2,1,-3,1}
  -4\, \HS_{-2,1,-2,-2}
  +4\, \HS_{-2,1,1,-3}
  +4\, \HS_{-2,2,-2,1}\nn\\
&&  -4\, \HS_{3,-2,1,1}
  -4\, \HS_{3,1,1,-2}
  +4\, \HS_{4,1,1,1}
  -8\, \HS_{-2,-2,1,1,1}
  -8\, \HS_{-2,1,1,-2,1})
\Big )
\label{finalwrapping}\,.
\eeqa

The analytic continuation of the full anomalous dimension \eqref{totalad} to the BFKL pole $M=-1+\omega$ is tedious but feasible. One finds \textit{full agreement} with the LO and NLO BFKL prediction! Also the double-logarithmic constraints are satisfied. This is a very non-trivial check of our findings and a convincing evidence that the first L\"uscher correction may be applied at the five-loop order without any modifications. In particular, the successful comparison with the BFKL equation indirectly supports the correctness of the $M=2$ result, i.e. the five-loop anomalous dimension of the Konishi operator, found in \cite{Bajnok:2009vm}.

This paper is structured as follows. In section \ref{sec:structure} we review the basic facts about the structure of the anomalous dimension of twist-two operators. In section \ref{sec:weak} we discuss the known constraints on the five-loop anomalous dimension. They provide a possibility of an \textit{independent} verification of our findings. We derive the  contribution of the asymptotic Bethe ansatz to the five-loop anomalous dimension in section \ref{sec:fiveloop}. At the end of the section, we also perform analytic continuations of the result relevant from the perspective of the constraints discussed in section \ref{sec:weak}. Finally, in section \ref{sec:wrapping} we employ the first L\"uscher correction to calculate the contribution of the wrapping diagrams and subsequently check it against the constraints discussed in section \ref{sec:weak}.
Some of the lengthy results and formulas are delegated to appendices. Aware of the inconvenience that typing in of our results may cause, we have decided to set up a webpage with \texttt{Mathematica} notebooks containing main results obtained in this article. It can be found under \href{http://thd.pnpi.spb.ru/~velizh/5loop/}{\texttt{http://thd.pnpi.spb.ru/\textasciitilde velizh/5loop/}}.

\section{The structure of the anomalous dimension}\label{sec:structure}
Twist-two operators belong to the $\mathfrak{sl}(2)$ sub-sector of the full theory. The highest-weight representative consists of two scalar fields $\fldZ$ and $M$ covariant derivatives $\fldD$
\begin{equation}
\label{twisttwo}
\Tr \left( \fldZ\, \fldD^\M\, \fldZ\,\right) + \ldots\, .
\end{equation}
Note that proper eigenstates of the dilatation operator are linear combinations of different distributions of the derivatives over the scalar fields. It is well-known that the anomalous dimension of these operators
governs the leading breaking of the Bjorken scaling. In the spin chain picture they are identified with the states of the non-compact $\alg{sl}(2)$
spin $=-\sfrac{1}{2}$ length-two Heisenberg magnet with $\M$ excitations. The cyclicity of the trace eliminates states with odd values of $\M$. For each even $\M$, on the other hand,
there is precisely one non-BPS state whose total scaling dimension is
\begin{equation}
\label{dimension}
\Delta=2+\M+\gamma(g)\, ,
\qquad {\rm with} \qquad
\gamma(g)=\sum_{\ell=1}^\infty  \gamma^{}_{2\ell}\,g^{2\ell}\, .
\end{equation}
Here, $\gamma(g)$ is the anomalous part of the dimension
depending on the coupling constant
\begin{equation}
\label{convention}
g^2=\frac{\lambda}{16\,\pi^2}\, ,
\end{equation}
and $\lambda=N\, \gym^2$ is the 't Hooft coupling constant. The anomalous
dimension $\gamma(g)$ may be
determined to the three-loop order $\Op(g^6)$ with help of the asymptotic
Bethe ansatz \cite{Staudacher:2004tk}. We will briefly discuss the $\alg{sl}(2)$ Bethe equations in section \ref{sec:fiveloop}.

Based on the observations made in \cite{Kotikov:2002ab, Kotikov:2004ss}, the authors of \cite{Kotikov:2007cy} have formulated the principle of maximal transcendentality. It assumes that at each order of the perturbation theory $\ell$ the anomalous dimension of twist-two operators is expressed through the generalised harmonic sums of the order $(2\ell-1)$, or through the products of zeta functions and harmonic sums for which the sum of the arguments of the zeta functions and the orders of the harmonic sums is equal to $(2\ell-1)$. The generalised harmonic sums are defined by the following recursive procedure (see \cite{Vermaseren:1998uu})
\beq \label{vhs}
S_a (M)=\sum^{M}_{j=1} \frac{(\mbox{sgn}(a))^{j}}{j^{\vert a\vert}}\, , \qquad
S_{a_1,\ldots,a_n}(M)=\sum^{M}_{j=1} \frac{(\mbox{sgn}(a_1))^{j}}{j^{\vert a_1\vert}}
\,S_{a_2,\ldots,a_n}(j)\, .
\eeq
The order $\ell$ of each sum $S_{a_1,\ldots,a_n}$ is given by the sum of the absolute values of its indices
\beq
\ell=\vert a_1 \vert +\ldots \vert a_n \vert\,,
\eeq
and the order of a product of harmonic sums is equal to the sum of the orders of its constituents. The canonical basis of the harmonic sums of $\ell$-th order is spanned by
\beqa \label{basis}
&&\left\{S_{a_{11}}, S_{a_{21},a_{22}},\, \ldots,\, S_{a_{\ell1},a_{\ell2},\ldots,a_{\ell\ell}}: a_{i j}
\in \mathbb{Z}\,,\right.\nonumber \\
&&\left.\qquad \qquad \ell=\vert a_{11} \vert=\vert a_{21} \vert+\vert a_{22} \vert=\ldots=\vert a_{\ell 1} \vert+\vert a_{\ell 2} \vert+\ldots+\vert a_{\ell\ell} \vert \right\}\,,
\eeqa
where the $\M$ dependence of the sums is implicit. Each $\ell$-th order product of harmonic sums can be decomposed in this basis. The one-, two- and three-loop anomalous dimensions are simple combinations of the sums \eqref{vhs} and may be found in \cite{Kotikov:2003fb,Kotikov:2004ss}. One can check \cite{Staudacher:2004tk} and even prove \cite{Kotikov:2008pv} that they coincide with the corresponding results of the ABE. This furnishes evidence of the validity of asymptotic integrability in the $\mathfrak{sl}(2)$ sector up to the $\Op(g^6)$ order in perturbation theory. In \cite{Kotikov:2007cy} a four-loop result, as predicted by the ABE, has been found. Also at this order the principle of maximal transcendentality applies, even when the correct wrapping contribution of \cite{Bajnok:2008qj} is added. We thus assume it to hold at the fifth order of perturbation theory as well.

A curious symmetry of the anomalous dimension of twist-two operators is the so-called reciprocity
\cite{Dokshitzer:2005bf,Dokshitzer:2006nm,Basso:2006nk}, which is a generalisation of the Gribov-Lipatov one-loop
reciprocity. In the reciprocity-respecting basis of harmonic sums the $\reciP$ function, which
according to the reciprocity relation \eqref{Pfunction} is closely related to the anomalous dimension,
takes much simpler form, as was shown for the first three orders of perturbation theory in
\cite{Dokshitzer:2006nm}. Recently, also the four-loop correction has been simplified in this manner
\cite{Beccaria:2009vt}. Reciprocity is a hidden symmetry of the ground states of twist-three operators
as well, at least up to five-loop order \cite{Beccaria:2007pb,Beccaria:2009eq}. In this article
we propose to determine the five-loop contribution to $\reciP(g)$ instead of $\gamma(g)$. Moreover, we
introduce an equivalent, but simpler, definition of the reciprocity-respecting basis at any loop
order. Eventually, the resulting set of harmonic sums that may contribute at any given order $\ell$ is
only a small subgroup of \eqref{basis}. Assuming reciprocity to be a symmetry at the fifth order of
perturbation theory, we will compute in section \ref{sec:fiveloop} the ABA part of the five-loop
$\reciP$ function along the methods proposed in \cite{Kotikov:2007cy}. It is tedious but
straightforward task to extract from it the corresponding five-loop anomalous dimension.
We present the result in Appendix \ref{sec:ABAresult}.

\section{Weak-coupling constraints}\label{sec:weak}
In this section we will discuss the known weak-coupling constraints on the five-loop anomalous dimension of twist-two operators. One class of constraints provides the NLO BFKL equation. A complementary set of constraints follow from the double-logarithmic behaviour of the amplitudes.

The relation between the anomalous dimension of twist-two operators and the Balitsky-Fadin-Kuraev-Lipatov (BFKL) equation \cite{Lipatov:1976zz,Kuraev:1977fs,Balitsky:1978ic} and its next-to-leading order (NLO) generalisation \cite{Fadin:1998py,Kotikov:2000pm}
 emerges upon analytic continuation of the
function $\gamma(g,M)$, and therefore at weak coupling of each of the $\gamma^{}_{2 \ell}(M)$,
to complex values of $\M$. This is straightforward in the one-loop case since
\beq
\gamma_{2}(M) = 8\,g^2\,S_1 (M) = 8\,g^2\, \left(\Psi(M+1)-\Psi(1)\right)\,,
\eeq
where $\Psi(x)=\frac{d}{dx}\,\log \Gamma(x)$ is the digamma function. The foundations for analytic continuation of more complicated harmonic sums have been laid in \cite{Kotikov:2005gr}.
At any loop order one expects
singularities at all \textit {negative integer} values of  $\M$.
The first in this series of singular points,
\begin{equation}
\label{omega}
M=-1+\omega\,,
\end{equation}
corresponds to the aforementioned pomeron. In the above formula $\omega$ should be considered infinitesimally small. The BFKL equation relates $\gamma(g)$ and $g$ in the vicinity
of the point $\M=-1+\omega$. It predicts that, if expanded in $g$, the $\ell$-loop anomalous dimension $\gamma_{2 \ell} (\omega)$ exhibits poles in $\omega$. Moreover, the residues and the order of the poles can be derived directly from the BFKL equation. The BFKL equation has been formulated up to the next-to-leading order (NLO) in the logarithmic expansion and determines the leading and next-to-leading poles of $\gamma_{2\ell}(\omega)$. Please refer to \cite{Kotikov:2007cy} for details and further explanations.
The NLO-BFKL equation for twist-two operators in the dimensional reduction scheme can be written as follows
\begin{equation}
\frac{\omega}{-4\,g^2} = \chi (\gamma )-g^2\,\delta (\gamma )\,,
\end{equation}
where
\begin{eqnarray}\label{gammanlo}
\chi (\gamma ) &=&
\Psi\left(-\frac{\gamma}{2}\right)+\Psi\left(1+\frac{\gamma}{2}\right)-
2\,\Psi\left(1\right)\, ,\\[4mm]
\delta (\gamma ) &=&4\,\chi ^{\,\prime \prime } (\gamma )
+6\,\zeta(3)+2\,\zeta(2)\,\chi (\gamma )+4\,\chi (\gamma )\,\chi ^{\,\prime} (\gamma )  \nonumber \\[2mm]
& & -\frac{\pi^3}{\sin \frac{\pi \gamma}{2}}- 4\,\Phi \left(-\frac{\gamma}{2}
\right) -4\,\Phi \left(1+\frac{\gamma}{2} \right)\,.
\end{eqnarray}
The function $\Phi (\gamma )$ is given by
\begin{eqnarray}
\Phi (\gamma ) =~\sum_{k=0}^{\infty }\frac{(-1)^{k}} {(k+\gamma)^2 }\biggl[\Psi
\left(k+\gamma +1\right)-\Psi (1)\biggr]. \label{9}
\end{eqnarray}
Upon using the expansion \eqref{dimension}, one easily determines the leading singularity structure
\begin{eqnarray}\label{NLOpoles}
 \gamma&=&\left(2+0\,\omega+\Op(\omega^2)\right)
\left(\frac{-4\,g^2}{\omega}\right) -\left(0+0\,\omega
+\Op(\omega^2)\right)\,\left(\frac{-4\,g^2}{\omega}\right)^2
\nonumber
\\
&&
+\left(0+\,\zeta(3)\,\omega +\Op(\omega^2) \right)\,\left(\frac{-4\,g^2}{\omega}\right)^3
-\left(4\,\zeta(3)+\frac{5}{4}\,\zeta(4)\,\omega +\Op(\omega^2)\right)\,\left(\frac{-4\,g^2}{\omega}\right)^4 \nonumber \\
&&
-\left(0+\bigg(2\,\zeta(2)\,\zeta(3)+16\,\zeta(5)\bigg)\,\omega+\Op(\omega^2)\right)\left(\frac{-4g^2}{\omega}\right)^5\pm \ldots .
\end{eqnarray}

Another class of constraints on the five-loop anomalous dimension provide the constraints following
from the double-logarithmic asymptotics of the scattering amplitudes. In QED and QCD this phenomenon
was extensively studied in~\cite{Gorshkov:1966ht,Gorshkov:1966hu}
and~\cite{Kirschner:1982xw,Kirschner:1982qf,Kirschner:1983di} (see also
\cite{Bartels:1995iu,Bartels:1996wc}). It amounts to summing up the leading terms $\sim (\alpha \ln
^2s)^n$ in all orders of perturbation theory. Please refer to \cite{Kotikov:2002ab, Kotikov:2000pm}
for the discussion of this limit in the case of the $\superN=4$ gauge theory. The resulting
constraints allow to determine the leading singularity at $M=-2+\omega$
\begin{eqnarray}\label{dlevenp}
\gamma&=&-\omega+\omega\, \sqrt{1-\frac{16 g^2}{\omega^2}}
\nonumber \\
&=&
2\,\frac{(-4\, g^2)}{\omega}
-2\,\frac{(-4\, g^2)^2}{\omega^3}
+4\,\frac{(-4\, g^2)^3}{\omega^5}
-10\,\frac{(-4\, g^2)^4}{\omega^7}+28\,\frac{(-4\, g^2)^5}{\omega^9}-\ldots\, .
\end{eqnarray}

\section{The Five-Loop Anomalous Dimension from Bethe Ansatz}
\label{sec:fiveloop}

The long-range asymptotic Bethe equations for the $\mathfrak{sl}(2)$ operators can be found directly from the full set of the asymptotic Bethe equations proposed in \cite{Beisert:2005fw,Beisert:2006ez}
\begin{equation}
\label{sl2eq}
\left(\frac{x^+_k}{x^-_k}\right)^L=\prod_{\substack{j=1\\j \neq k}}^\M
\frac{x_k^--x_j^+}{x_k^+-x_j^-}\,
\frac{1-g^2/x_k^+x_j^-}{1-g^2/x_k^-x_j^+}\,
\exp\left(2\,i\,\theta(u_k,u_j)\right),
\qquad
\prod_{k=1}^M \frac{x^+_k}{x^-_k}=1\, .
\end{equation}
There are $M$ equations for $k=1,\ldots,\M$ which need to be solved for the Bethe roots
$u_k$. The variables $x^{\pm}_k$ are related to $u_k$ through Zhukovsky map
\begin{equation}\label{definition x}
x_k^{\pm}=x(u_k^\pm)\, ,
\qquad
u^\pm=u\pm\tfrac{i}{2}\, ,
\qquad
x(u)=\frac{u}{2}\left(1+\sqrt{1-4\,\frac{g^2}{u^2}}\right).
\end{equation}
The function $\theta(u,v)$ is the celebrated dressing phase and has been
 conjectured in
\cite{Beisert:2006ez}. To the fifth order in perturbation theory it is sufficient to write
\begin{equation}
\label{4loopphase}
\theta(u_k,u_j) =\left(4\,
\zeta(3)\,g^6 -40\, \zeta(5) g^8 \right) \big(q_2(u_k)\,q_3(u_j)-q_3(u_k)\,q_2(u_j)\big)
+\Op(g^{10})\, ,
\end{equation}
where $q_r(u)$ are the eigenvalues of the conserved
magnon charges,  see \cite{Beisert:2005fw}. Note that  \eqref{sl2eq} has in general many solutions corresponding to different eigenstates of the dilatation operator. Once a solution has been found, the corresponding asymptotic
all-loop anomalous dimension is given by
\begin{equation}
\label{dim}
\gaba(g)=2\, g^2\, \sum^{\M}_{k=1}
\left(\frac{i}{x^{+}_k}-\frac{i}{x^{-}_k}\right) .
\end{equation}
It is related to the \textit{full} anomalous dimension through
\beq
\gamma (g) = \gaba (g)+\Delta_{w}(g)\,.
\eeq
The second term is the contribution of the aforementioned wrapping diagrams and for a state of length $L$ it should be taken into account starting from the order $\Op(g^{2L+4})$. In this section we
 determine $\gaba(g)$ at the fifth order in perturbation theory.

Assuming the transcendentality principle discussed in section \ref{sec:structure} together with several other observations related to the nested harmonic sums that cannot contribute, \textit{cf}. \cite{Kotikov:2007cy}, the basis \eqref{basis} reduces to 1500 sums, from which 108 may be attributed to the contribution of the dressing phase. Thus, following the method discussed in \cite{Kotikov:2007cy}, one would have to expand $\gaba_{10}$ in this basis and determine the coefficients by means of solving the $\mathfrak{sl}(2)$ asymptotic Bethe equations \eqref{sl2eq} to the precision which would allow to \textit{rationalise} $\gaba_{10}(M)$ for $M=1, \ldots , 1392$\footnote{Although the states for odd values of $M$ do not exist, one can still find the corresponding solutions of the Bethe equations. They, however, do not satisfy the momentum constraint. It was argued in \cite{Kotikov:2007cy} that $\gaba_{2l}(M)$ may be consistently continued with respect to the ABE to odd values of $M$.}. Despite the fact that the one-loop solution is known \cite{Kotikov:2007cy} and that the derivation of the higher-order corrections is merely a linear problem, it is beyond the computational threshold to determine $\gaba_{10}$ for such high values of $M$.
Fortunately, thanks to the reciprocity
\cite{Dokshitzer:2005bf,Dokshitzer:2006nm}
the basis may be further reduced to 256 sums, which renders the computation feasible! The asymptotic anomalous dimension, however, is not the right quantity to look at if one wants to make use of this simplification. Indeed, let us define a function $\reciP^{\mbox{\tiny ABA}}(N)$ \cite{Basso:2006nk}, such that
\beq \label{Pfunction}
\gaba(M) = \reciP^{\mbox{\tiny ABA}} \left(M+\frac{1}{2} \gaba(M) \right)\,.
\eeq
If not supplemented by further constraints on the structure of $\reciP^{\mbox{\tiny ABA}}(N)$, this relation is trivial. The reciprocity constrains $\reciP^{\mbox{\tiny ABA}}(N)$ to
\beq \label{PMexpansion}
\reciP^{\mbox{\tiny ABA}}(M) = \sum_{\ell \geq 0} \frac{a_{\ell}(\log J^2)}{J^{2 \ell}}, \quad J^2=M(M+1) \gg 1\,,
\eeq
with the functions $a_{\ell}(N)$ being polynomials. Upon substituting the perturbative expansion \eqref{dimension}, one finds to the five-loop order
\beq
\reciP^{\mbox{\tiny ABA}}(M)=g^2\,\reciP_2(M)+g^4\,\reciP_4(M)+g^6\,\reciP_6(M)+g^8\,\reciP_8(M)+g^{10}\,\reciP_{10}(M)+\ldots\,,
\eeq
with the coefficients $\reciP_{2i} (M), \ i=1,\ldots, 5$ taking the following form
\beqa
 \reciP_2(M) &=&
\gamma _2 \,,
\label{reciP2}\\
 \reciP_4(M) &=&
\gamma _4
-\frac{1}{2} \dotgamma_2 \gamma _2\,,
\label{reciP4}\\
 \reciP_6(M) &=&
\gamma _6
+\frac{1}{4} \gamma _2 \dotgamma_2^2
-\frac{1}{2} \gamma _4 \dotgamma_2
+\frac{1}{8} \ddotgamma_2 \gamma _2^2
-\frac{1}{2} \dotgamma_4 \gamma _2\,,
\label{reciP6}\\
 \reciP^{\textrm{rational}}_8(M) &=&
\gamma^{\textrm{rational}} _8
-\frac{1}{8} \gamma _2 \dotgamma_2^3
+\frac{1}{4} \gamma _4 \dotgamma_2^2
-\frac{3}{16} \ddotgamma_2 \gamma _2^2 \dotgamma_2
+\frac{1}{2} \dotgamma_4 \gamma _2 \dotgamma_2
-\frac{1}{2} \gamma _6 \dotgamma_2
-\frac{1}{48} \dddotgamma_2 \gamma_2^3
\nonumber\\&&
+\frac{1}{8} \ddotgamma_4 \gamma _2^2-\frac{1}{2} \dotgamma_6 \gamma_2
-\frac{1}{2} \dotgamma_4 \gamma_4
+\frac{1}{4} \ddotgamma_2 \gamma_2 \gamma_4\,,
\label{reciP8rat}\\
 \reciP^{\textrm{rational}}_{10}(M) &=&
\gamma^{\textrm{rational}} _{10}
+\frac{1}{16} \gamma _2 \dotgamma_2^4
-\frac{1}{8} \gamma _4 \dotgamma_2^3
+\frac{3}{16} \ddotgamma_2 \gamma _2^2 \dotgamma_2^2
-\frac{3}{8} \dotgamma_4 \gamma_2 \dotgamma_2^2
+\frac{1}{4} \gamma _6 \dotgamma_2^2
\nonumber\\&&
+\frac{1}{24} \dddotgamma_2 \gamma_2^3 \dotgamma_2-\frac{3}{16} \ddotgamma_4 \gamma_2^2 \dotgamma_2
+\frac{1}{2}  \dotgamma_6 \gamma_2 \dotgamma_2
+\frac{1}{2} \dotgamma_4 \gamma_4 \dotgamma_2
-\frac{3}{8} \ddotgamma_2 \gamma _2 \gamma _4\dotgamma_2
-\frac{1}{2} \gamma^{\textrm{rational}}_8 \dotgamma_2
\nonumber\\&&
+\frac{1}{384} \ddddotgamma_2 \gamma _2^4+\frac{1}{32} \ddotgamma_2^2 \gamma _2^3
-\frac{1}{48} \dddotgamma_4 \gamma _2^3
+\frac{1}{8} \ddotgamma_6 \gamma _2^2
-\frac{3}{16} \ddotgamma_2 \dotgamma_4 \gamma_2^2
+\frac{1}{8} \ddotgamma_2 \gamma_4^2
+\frac{1}{4} \dotgamma_4^2 \gamma_2
\nonumber\\&&
-\frac{1}{2} \dotgamma^{\textrm{rational}}_8\gamma _2 -\frac{1}{16} \dddotgamma_2 \gamma _2^2 \gamma_4
-\frac{1}{2} \dotgamma_6 \gamma _4
+\frac{1}{4} \ddotgamma_4 \gamma_2 \gamma_4
-\frac{1}{2} \dotgamma_4 \gamma _6
+\frac{1}{4} \ddotgamma_2 \gamma _2 \gamma _6\,,
\label{reciP10rat}\\
\reciP^{\zeta(3)}_8(M)&=&\gamma^{\zeta(3)}_8\,,
\label{reciP8z3}\\
 \reciP_{10}^{\zeta(3)}(M)&=&
\gamma_{10}^{\zeta(3)}
-\frac{1}{2} \gamma_8^{\zeta(3)} \dotgamma_2
-\frac{1}{2} \dotgamma_8^{\zeta(3)} \gamma_2
\,,
\label{reciP10z3}\\
 \reciP_{10}^{\zeta(5)}(M)&=&
\gamma_{10}^{\zeta(5)}\,.
\label{reciP10z5}
\eeqa
Here, dots over $\gamma$ indicate derivatives of the harmonic sums with
respect to their indices (see \cite{Dokshitzer:2005bf,Dokshitzer:2006nm}),
while $\gamma^{\textrm{rational}}_i$, $\gamma^{\zeta(3)}_i$ and
$\gamma^{\zeta(5)}_i$ denote respectively the rational, $\zeta(3)$ and
$\zeta(5)$ parts of the anomalous dimension $\gamma_i$. Note that
$\reciP_{2i}(M)$ depend linearly on $\gaba_{2i}$. The functions $\reciP_8$
may be found in \cite{Beccaria:2009vt}.
Due to the property \eqref{PMexpansion}, only certain combinations of harmonic
sums may contribute to $\reciP_{2i}$.
We discuss the basis of harmonic sums respecting \eqref{PMexpansion} in \ref{app:rrsums}. It turns out that at the five-loop order only 256 of these sums need to be taken into account.
With computational effort of more then 400 hours of computer time,
we have found the following result for $\reciP_{10}(M)$
\begin{table}
\beqa
\frac{\reciP^{\textrm{rational}}_{10}}{128}&=&
-5 \, \HBS_{2,2,5}
-\, \HBS_{2,6,1}
+19 \, \HBS_{3,1,5}
-20 \, \HBS_{3,2,4}
+21 \, \HBS_{4,1,4}
-24 \, \HBS_{4,2,3}
+25 \, \HBS_{5,1,3}\nonumber\\ & &
-18 \, \HBS_{5,2,2}
+7 \, \HBS_{6,1,2}
-4 \, \HBS_{6,2,1}
-2 \, \HBS_{1,1,2,5}
+2 \, \HBS_{1,1,6,1}
-2 \, \HBS_{1,2,1,5}
-\, \HBS_{1,2,2,4}\nonumber\\ & &
+\, \HBS_{1,2,3,3}
+\, \HBS_{1,2,4,2}
-6 \, \HBS_{1,2,5,1}
+23 \, \HBS_{1,3,1,4}
-24 \, \HBS_{1,3,2,3}
-\, \HBS_{1,3,4,1}
+23 \, \HBS_{1,4,1,3}\nonumber\\ & &
-20 \, \HBS_{1,4,2,2}
-\, \HBS_{1,4,3,1}
+13 \, \HBS_{1,5,1,2}
-12 \, \HBS_{1,5,2,1}
+6 \, \HBS_{1,6,1,1}
-2 \, \HBS_{2,1,1,5}
+5 \, \HBS_{2,1,2,4}\nonumber\\ & &
+\, \HBS_{2,1,3,3}
+\, \HBS_{2,1,4,2}
-5 \, \HBS_{2,1,5,1}
-16 \, \HBS_{2,2,1,4}
+17 \, \HBS_{2,2,2,3}
-2 \, \HBS_{2,2,3,2}
+14 \, \HBS_{2,2,4,1}\nonumber\\ & &
-29 \, \HBS_{2,3,1,3}
+25 \, \HBS_{2,3,2,2}
+4 \, \HBS_{2,3,3,1}
-19 \, \HBS_{2,4,1,2}
+20 \, \HBS_{2,4,2,1}
-12 \, \HBS_{2,5,1,1}
+20 \, \HBS_{3,1,1,4}\nonumber\\ & &
-22 \, \HBS_{3,1,2,3}
-8 \, \HBS_{3,1,3,2}
+6 \, \HBS_{3,1,4,1}
-26 \, \HBS_{3,2,1,3}
+36 \, \HBS_{3,2,2,2}
-5 \, \HBS_{3,2,3,1}
-6 \, \HBS_{3,3,1,2}\nonumber\\ & &
+5 \, \HBS_{3,3,2,1}
-2 \, \HBS_{3,4,1,1}
+22 \, \HBS_{4,1,1,3}
-24 \, \HBS_{4,1,2,2}
+6 \, \HBS_{4,1,3,1}
-18 \, \HBS_{4,2,1,2}
+18 \, \HBS_{4,2,2,1}\nonumber\\ & &
-2 \, \HBS_{4,3,1,1}
+14 \, \HBS_{5,1,1,2}
-10 \, \HBS_{5,1,2,1}
-14 \, \HBS_{5,2,1,1}
+8 \, \HBS_{6,1,1,1}
+4 \, \HBS_{1,1,1,1,5}
-6 \, \HBS_{1,1,1,2,4}\nonumber\\ & &
-2 \, \HBS_{1,1,1,3,3}
-2 \, \HBS_{1,1,1,4,2}
+6 \, \HBS_{1,1,1,5,1}
-4 \, \HBS_{1,1,2,1,4}
+4 \, \HBS_{1,1,2,2,3}
+5 \, \HBS_{1,1,2,3,2}\nonumber\\ & &
-13 \, \HBS_{1,1,2,4,1}
+24 \, \HBS_{1,1,3,1,3}
-20 \, \HBS_{1,1,3,2,2}
-5 \, \HBS_{1,1,3,3,1}
+16 \, \HBS_{1,1,4,1,2}
-19 \, \HBS_{1,1,4,2,1}\nonumber\\ & &
+12 \, \HBS_{1,1,5,1,1}
-4 \, \HBS_{1,2,1,1,4}
+7 \, \HBS_{1,2,1,2,3}
+4 \, \HBS_{1,2,1,3,2}
-8 \, \HBS_{1,2,1,4,1}
-19 \, \HBS_{1,2,2,1,3}\nonumber\\ & &
+9 \, \HBS_{1,2,2,2,2}
+24 \, \HBS_{1,2,2,3,1}
-22 \, \HBS_{1,2,3,1,2}
+31 \, \HBS_{1,2,3,2,1}
-22 \, \HBS_{1,2,4,1,1}
+22 \, \HBS_{1,3,1,1,3}\nonumber\\ & &
-24 \, \HBS_{1,3,1,2,2}
+6 \, \HBS_{1,3,1,3,1}
-20 \, \HBS_{1,3,2,1,2}
+23 \, \HBS_{1,3,2,2,1}
-6 \, \HBS_{1,3,3,1,1}
+16 \, \HBS_{1,4,1,1,2}\nonumber\\ & &
-11 \, \HBS_{1,4,1,2,1}
-22 \, \HBS_{1,4,2,1,1}
+16 \, \HBS_{1,5,1,1,1}
-4 \, \HBS_{2,1,1,1,4}
+7 \, \HBS_{2,1,1,2,3}
+4 \, \HBS_{2,1,1,3,2}\nonumber\\ & &
-8 \, \HBS_{2,1,1,4,1}
+3 \, \HBS_{2,1,2,1,3}
-11 \, \HBS_{2,1,2,2,2}
+14 \, \HBS_{2,1,2,3,1}
-9 \, \HBS_{2,1,3,1,2}
+16 \, \HBS_{2,1,3,2,1}\nonumber\\ & &
-12 \, \HBS_{2,1,4,1,1}
-16 \, \HBS_{2,2,1,1,3}
+13 \, \HBS_{2,2,1,2,2}
+5 \, \HBS_{2,2,1,3,1}
+28 \, \HBS_{2,2,2,1,2}
-67 \, \HBS_{2,2,2,2,1}\nonumber\\ & &
+32 \, \HBS_{2,2,3,1,1}
-22 \, \HBS_{2,3,1,1,2}
+15 \, \HBS_{2,3,1,2,1}
+31 \, \HBS_{2,3,2,1,1}
-23 \, \HBS_{2,4,1,1,1}
+20 \, \HBS_{3,1,1,1,3}\nonumber\\ & &
-22 \, \HBS_{3,1,1,2,2}
+6 \, \HBS_{3,1,1,3,1}
-8 \, \HBS_{3,1,2,1,2}
+2 \, \HBS_{3,1,2,2,1}
+6 \, \HBS_{3,1,3,1,1}
-20 \, \HBS_{3,2,1,1,2}\nonumber\\ & &
+15 \, \HBS_{3,2,1,2,1}
+25 \, \HBS_{3,2,2,1,1}
-7 \, \HBS_{3,3,1,1,1}
+16 \, \HBS_{4,1,1,1,2}
-11 \, \HBS_{4,1,1,2,1}
-11 \, \HBS_{4,1,2,1,1}\nonumber\\ & &
-23 \, \HBS_{4,2,1,1,1}
+16 \, \HBS_{5,1,1,1,1}
-4 \, \HBS_{1,1,1,1,2,3}
-4 \, \HBS_{1,1,1,1,3,2}
+8 \, \HBS_{1,1,1,1,4,1}
+4 \, \HBS_{1,1,1,2,2,2}\nonumber\\ & &
-13 \, \HBS_{1,1,1,2,3,1}
+18 \, \HBS_{1,1,1,3,1,2}
-25 \, \HBS_{1,1,1,3,2,1}
+16 \, \HBS_{1,1,1,4,1,1}
+3 \, \HBS_{1,1,2,1,2,2}\nonumber\\ & &
-5 \, \HBS_{1,1,2,1,3,1}
-20 \, \HBS_{1,1,2,2,1,2}
+53 \, \HBS_{1,1,2,2,2,1}
-33 \, \HBS_{1,1,2,3,1,1}
+18 \, \HBS_{1,1,3,1,1,2}\nonumber\\ & &
-11 \, \HBS_{1,1,3,1,2,1}
-25 \, \HBS_{1,1,3,2,1,1}
+20 \, \HBS_{1,1,4,1,1,1}
+3 \, \HBS_{1,2,1,1,2,2}
-5 \, \HBS_{1,2,1,1,3,1}\nonumber\\ & &
+14 \, \HBS_{1,2,1,2,2,1}
-17 \, \HBS_{1,2,1,3,1,1}
-18 \, \HBS_{1,2,2,1,1,2}
+11 \, \HBS_{1,2,2,1,2,1}
+52 \, \HBS_{1,2,2,2,1,1}\nonumber\\ & &
-27 \, \HBS_{1,2,3,1,1,1}
+16 \, \HBS_{1,3,1,1,1,2}
-11 \, \HBS_{1,3,1,1,2,1}
-11 \, \HBS_{1,3,1,2,1,1}
-27 \, \HBS_{1,3,2,1,1,1}\nonumber\\ & &
+20 \, \HBS_{1,4,1,1,1,1}
+3 \, \HBS_{2,1,1,1,2,2}
-5 \, \HBS_{2,1,1,1,3,1}
+14 \, \HBS_{2,1,1,2,2,1}
-17 \, \HBS_{2,1,1,3,1,1}\nonumber\\ & &
+16 \, \HBS_{2,1,2,2,1,1}
-11 \, \HBS_{2,1,3,1,1,1}
-16 \, \HBS_{2,2,1,1,1,2}
+11 \, \HBS_{2,2,1,1,2,1}
+11 \, \HBS_{2,2,1,2,1,1}\nonumber\\ & &
+45 \, \HBS_{2,2,2,1,1,1}
-27 \, \HBS_{2,3,1,1,1,1}
+16 \, \HBS_{3,1,1,1,1,2}
-11 \, \HBS_{3,1,1,1,2,1}
-11 \, \HBS_{3,1,1,2,1,1}\nonumber\\ & &
-11 \, \HBS_{3,1,2,1,1,1}
-27 \, \HBS_{3,2,1,1,1,1}
+20 \, \HBS_{4,1,1,1,1,1}
-16 \, \HBS_{1,1,1,1,2,2,1}
+24 \, \HBS_{1,1,1,1,3,1,1}\nonumber\\ & &
-28 \, \HBS_{1,1,1,2,2,1,1}
+20 \, \HBS_{1,1,1,3,1,1,1}
-20 \, \HBS_{1,1,2,2,1,1,1}
+20 \, \HBS_{1,1,3,1,1,1,1}\nonumber\\ & &
-20 \, \HBS_{1,2,2,1,1,1,1}
+20 \, \HBS_{1,3,1,1,1,1,1}
-20 \, \HBS_{2,2,1,1,1,1,1}
+20 \, \HBS_{3,1,1,1,1,1,1}\,.
\eeqa
\caption{The five-loop function $\reciP_{10} (M)$.} \label{pfiveloop}
\end{table}
\newpage
\noindent The functions $\reciP^{\zeta(3)}_{10}$ and $\reciP^{\zeta(5)}_{10}$ are easy to determine\footnote{The $\zeta(3)$ and $\zeta(5)$ parts of the  five-loop ABA contribution to the anomalous dimension have already been found in \cite{Beccaria:2009rw}.}\footnote{Curiously, it seems that at the $\ell$-loop order the highest transcendentality contribution of the dressing phase to the anomalous dimension may be found exactly: $-64\, \beta^{\ell}_{2, 3}\, S_1\,( S_3 - S_{-3} + 2 \,S_{-2, 1})$.}
\beqa
\frac{\reciP_{10}^{\zeta(3)}}{256}&=&
3 \HBS_{1,5}
-4 \HBS_{2,4}
-\HBS_{3,3}
-\HBS_{4,2}
+3 \HBS_{5,1}
+2 \HBS_{1,1,4}
-4 \HBS_{1,2,3}
-2 \HBS_{1,3,2}
+2 \HBS_{1,4,1}\nn\\
&&-\HBS_{2,1,3}
+5 \HBS_{2,2,2}
-2 \HBS_{2,3,1}
-2 \HBS_{3,2,1}
+2 \HBS_{4,1,1}
-3 \HBS_{1,1,2,2}
+\HBS_{1,1,3,1}
-\HBS_{1,2,1,2}\nn\\
&&+2 \HBS_{1,3,1,1}
-\HBS_{2,1,1,2}
+\HBS_{2,1,2,1}
+\HBS_{3,1,1,1}\,,\label{ABAz3}\\
\frac{\reciP_{10}^{\zeta(5)}}{640}&=&\HBS_1 (\HBS_{2,1}-\HBS_3)\,.\label{ABAz5}
\eeqa
The resulting asymptotic anomalous dimension,
\textit{cf.} table \ref{ABAresult} in Appendix \ref{sec:ABAresult},
may now be analytically continued to the BFKL pole $\M=-1+\omega$.
With help of the \texttt{SUMMER}~\cite{Vermaseren:1998uu} and \texttt{HARMPOL}~\cite{Remiddi:1999ew}
packages for \texttt{FORM}~\cite{Vermaseren:2000nd} and the \texttt{HPL} package~\cite{Maitre:2005uu}
for \texttt{Mathematica}, we have found
\beqa\nn
\gaba&=&\left(\frac{-4\,g^2}{\omega}\right)
\bigg(2+0\omega
\bigg)
+\,\left(\frac{-4\,g^2}{\omega}\right)^2
\bigg(0+0\omega
\bigg)
+\,\left(\frac{-4\,g^2}{\omega}\right)^3
\bigg(0+\zeta(3)\omega
\bigg)
\nn\\&&
+\,\frac{(-4\,g^2)^4}{\omega^7}
\bigg(-2
+\frac{4 \pi ^2}{3} \omega^2
-13 \zeta(3) \omega^3
-\frac{8 \pi ^4}{45} \omega^4
\bigg)
+\frac{(-4g^2)^5}{\omega^9}\times
\nn\\&&
\bigg(2
-\frac{7 \pi ^2}{3} \omega^2
-\frac{39 \zeta(3)}{2} \omega^3
+\frac{361 \pi ^4}{240} \omega^4
+\frac{618\, \zeta(2) \,\zeta(3)-1199 \, \zeta(5)}{8} \omega^5
\bigg)
\nn\\&&
\pm \ldots\,, \label{killerexpansion}
\eeqa
where we have also restated the known results at lower orders. One observes a {\it maximal violation} of
the NLO BFKL prediction \eqref{gammanlo} at the four- and five-loop order; the leading singularity in $\omega$ should in both cases
 be a pole of order four. Instead, we
find the leading poles to be of order seven at the four-loop order and of order nine at the next order, thus providing three constraints on the wrapping contribution at the four-loop order and four constraints at five-loops. This counting takes into account the fact that the cancelation of the highest pole in the ABA result is intimately related to the cancelation of the next, lower pole. Please note that after adding the four-loop wrapping correction found in
\cite{Bajnok:2008qj} the agreement between \eqref{NLOpoles} and \eqref{killerexpansion} is restored at the four-loop order!

\section{Calculation of the wrapping correction} \label{sec:wrapping}

In this section we calculate the wrapping correction by evaluating the first L\"uscher correction at weak-coupling along the lines advocated in \cite{Bajnok:2008bm}, \cite{Bajnok:2008qj} and \cite{Bajnok:2009vm}. Note that the second L\"uscher correction is not expected to contribute before the order $\Op(g^{12})$. Please refer to \cite{Bajnok:2009vm} for further discussion.

The total wrapping correction consists of two terms
\beq \label{wrappingtotal}
\Delta_{w}=\Delta_{w}^F+\Delta_{w}^{\textrm{ABA}}\,.
\eeq
The first term, the so-called F-term,  reflects the finite-size correction to the dispersion formula, while the second term accounts for the shifts in the magnon rapidities induced by the interactions with the virtual particle.

Following the notation in \cite{Bajnok:2009vm}, the F-term integral can be written as
\begin{eqnarray} \label{fterm}
\Delta^F_{w}  &=&
-\sum_{Q=1}^{\infty}\int_{-\infty}^{\infty}\!\frac{dq}{2\pi}\Big(\frac{z^{-}}{z^
{+}}\Big)^{2}\sca(M,q,Q) \dress(M,q,Q) \mat(M,q,Q)^2 \\ \nonumber
&\equiv &-\sum_{Q=1}^{\infty}\int_{-\infty}^{\infty}\!\frac{dq}{2\pi}\intd(M,q)\,,
\end{eqnarray}
where $\intd(M,q)$ has the perturbative expansion of the form
\begin{equation} \label{Yexpansion}
\intd(M,q)=\intd^{(8,0)}(M,q)g^8+\Big(\intd^{(10,0)}(M,q)+\intd^{(8,2)}(M,q)\Big) g^{10}+\Op(g^{12})\,.
\end{equation}
The five-loop integrand coming from the F-term has two components. The first one, $Y_Q^{(10,0)}$, follows from expanding the integrand \eqref{fterm} to the five-loop order and inserting the one-loop Bethe roots $u_k=u^0_k + \Op(g^2)$. The second term, $Y^{(8,2)}_Q$ accounts for the two-loop corrections to the rapidities of the magnons. Clearly, while calculating $Y^{(8,2)}_Q$, it is sufficient to determine $Y^8_Q$ as a function of the parameters $u_k$ and subsequently put $u_k= u_k^0+u_k^2 g^2 +\Op(g^4)$
\beq
Y^{8}_Q(u_1,\dots,u_M,q)=Y^{(8,0)}_Q (M,q)+g^2\,Y^{(8,2)}_Q (M,q)+\Op(g^4)\,.
\eeq

In order to calculate $\intd^{(10,0)}$ we factor out the leading integrand and rewrite it as a sum of the matrix part, the scalar part, the exponential term and the dressing factor
\beqa \nn
\intd^{(10,0)}(M,q) &=&
\intd^{(8,0)}(M,q)\bigg[2\frac{S_{\boxplus}^{(4)}(M,q,Q)}{S_{\boxplus}^{(2)}(M,q,Q)
}+\frac{S_{0}^{(2)}(M,q,Q)}{S_{0}^{(0)}(M,q,Q)}+
\frac{\ex^{(6)}(q,Q)}{\ex^{(4)}(q,Q)}\\
&&+\dress^{(2)}(M,q,Q)\bigg]\,.
\eeqa
This section is structured as follows. In sub-sections \ref{sec:Y10} and \ref{sec:Y82} we determine $Y^{(10,0)}_Q$ and $Y^{(8,2)}_Q$. In sub-section \ref{sec:ABAmod} we will discuss the modification of the quantization condition, leading to $\Delta^{\textrm{ABA}}_w$.
\subsection{The four-loop integrand $Y^{(8,0)}_Q$}
Before proceeding to the five-loop order we recall the reader the form of $Y^{(8,0)}_Q(M,q)$, as found in \cite{Bajnok:2008qj}
\begin{equation}
\label{e.integral}
\intd^{(8,0)}(M,q)=64\, S_{1}^{2}\,
\frac{T_M(q,Q)^{2}}{R_M(q,Q)}\frac{16}{(q^{2}+Q^{2})^{2}}\,.
\end{equation}
In this formula
\begin{equation}
R^{(0)}_M(q,Q)=\Baxter(\frac{q - i(-1 + Q)}{2})\Baxter(\frac{q + i(-1 + Q)}{2})\Baxter(\frac{q - i(1 + Q)}{2})\Baxter(\frac{q + i(1 + Q)}{2})\,,
\end{equation}
and
\begin{equation}
T^{(0)}_M(q,Q)=\sum_{j=0}^{Q-1}\Big( \frac{1}{2 j - i q - Q} -(-1)^M \frac{1}{2 (j + 1) - i q - Q} \Big) \Baxter\big(\frac{q - i (Q - 1)}{2}  + i j\big)\,.
\end{equation}
The one-loop Baxter function, $\Baxter$, is a hypergeometric orthogonal polynomial \cite{Dippel}
\beq
\Baxter(u)={}_3 F_2\left(\left. \begin{array}{c}
-M, \ M+1,\ \frac{1}{2}+iu \\
1,\  1 \end{array}
\right| 1\right) \,.
\eeq

\subsection{Derivation of $Y^{(10,0)}$} \label{sec:Y10}
By comparing \eqref{fterm} and \eqref{Yexpansion}, one concludes that the calculation of $Y^{(10,0)}(M,q)$ amounts to expanding $(z^-/z^+)^2$, $\sca(M,q,Q)$, $\dress(M,q,Q)$ and $\mat(M,q,Q)$ to the next-to-leading order. This is two-loop order for the scalar and the dressing part, whilst the matrix part vanishes for $g=0$ and needs to be evaluated to three-loop order. On the other hand, the exponential part $(z^-/z^+)^2$ vanishes at the first two orders in $g^2$ and thus needs to be expanded to the order $\Op(g^8)$. Owing to the integration in \eqref{fterm}, we symmetrise all results with respect to $q$. The expressions obtained are valid for all \textit{positive integer} values of $M$.

\subsubsection*{Scalar Part}
Similarly to the case of the Konishi operator \cite{Bajnok:2009vm} the scalar part can be divided into two parts
\beq
\sca(M,q,Q)=\sca^{(0)}(M,q,Q)+g^2\,\Big(S_{0rat}^{(2)}(M,q,Q)+S_{0\psi}^{(2)}(M,q,Q)\Big)+\Op(g^4)\,.
\eeq
 The first part contains solely rational functions and may be expressed through one-loop Baxter function and its derivative
\beqa \nn
\frac{S_{0rat}^{(2)}(M,q,Q)}{\sca^{(0)}(M,q,Q)}&=&\frac{4\,q}{q^2+Q^2}\Big( \frac{\DBaxter(\frac{-q - i(1 + Q)}{2})}{\Baxter (\frac{-q - i(1 + Q)}{2})} - \frac{ \DBaxter (\frac{q - i(1 + Q)}{2})}{ \Baxter (\frac{(q - i(1 + Q)}{2})}\Big) - 8\,S_{-2}(M) \\
&&-
 \frac{32\,Q}{q^2 + Q^2}S_{1}(M)\,.
\eeqa
The second part consists of polygamma functions and depends on $M$ only through the harmonic number
\begin{equation}
\frac{S_{0\psi}^{(2)}(M,q,Q)}{\sca^{(0)}(q,Q,u)}=4\,S_1(M) \Big(
\psi\big( \frac{-i q - Q}{2}\big) -
  \psi\big(\frac{-i q + Q}{2} \big)+\psi\big( \frac{i q - Q}{2}\big) -
  \psi\big(\frac{i q + Q}{2} \big) \Big)\,.
\end{equation}

\subsubsection*{Dressing Part}
In contradistinction to the physical kinematics, the dressing factor governing the scattering of virtual particles on the physical magnons is trivial
 only at the leading order, i.e. $\dress^{(0)}=1$, \textit{cf.} \cite{Bajnok:2009vm}. At the next-to-leading order one finds
 \beqa \nn
\dress^{(2)}(M,q,Q)&=&-4\,S_1(M)\Big(4\gamma_E +  \psi\big(
 \frac{-i q - Q}{2}\big) +
    \psi\big( \frac{-i q + Q}{2} \big)\\
&& +\psi\big(
 \frac{i q - Q}{2}\big)  +
    \psi\big( \frac{i q + Q}{2} \big) \Big)\,.
\eeqa
When combined with the non-rational part of the scalar factor $S_{0\psi}^{(2)}$, they simplify further to
\begin{equation} \label{polygammas}
\frac{S_{0\psi}^{(2)}(M,q,Q)}{\sca^{(0)}(q,Q,u)}+\dress^{(2)}(M,q,Q)=-8\,S_1(M)\Big(2\gamma_E +
    \psi\big( \frac{-i q + Q}{2} \big)+
    \psi\big( \frac{i q + Q}{2} \big) \Big)\,.
\end{equation}
This is the only part of the integrand containing polygamma functions.

\subsubsection*{Exponential Part}
The exponential part does not depend on $M$ and is straightforward to determine
\begin{equation}
\frac{\ex^{(6)}(q,Q)}{\ex^{(4)}(q,Q)}=-\frac{16}{q^2 + Q^2} \,.
\end{equation}

\subsubsection*{The matrix part}
The calculation of the matrix part amounts to calculating the supertrace
\beq \label{matrixstr}
\textrm{str} \bigg(\prod^M_{k=1} S^{\textrm{\tiny matrix}}_Q(q,u_k)\bigg)\,
\eeq
to the three loop-order. This may be systematically achieved by introducing the matrix $G$ that diagonalizes each copy $\hat{S}_{\mathfrak{su}(2|2),Q}(q,u_k)$ of the tensor product
\beq \label{tensorproduct}
S^{\textrm{\tiny matrix}}_{Q}(q,u_k)=\hat{S}_{\mathfrak{su}(2|2),Q}(q,u_k) \otimes\hat{S}_{\mathfrak{su}(2|2),Q}(q,u_k)\,.
\eeq
Explicitly,
\beq
G \,\hat{S}_{\mathfrak{su}(2|2),Q}(q,u_k) \, G^{-1}=\hat{S}^{\textrm{\tiny diag}}_{\mathfrak{su}(2|2),Q}(q,u_k)\,.
\eeq
The key feature of $G$ is that to the order $\Op(g^4)$ it \textit{does not} depend on the parameters $u_k$, i.e. $G=G(q,Q)$. The diagonal matrix $S^{\textrm{\tiny diag}}_Q(q,u_k)$ has the following structure
\beqa \nn
&&\bigg\{\hat{S}^{\textrm{\tiny diag}}_{\mathfrak{su}(2|2),Q}(q,u_k)\bigg\}_{ii}=B_1(i,u_k)\theta(Q-i)+B_2(i-Q,u_k)\theta(2Q-i)\theta(i-Q)\\ \label{smatrixstructure}
&&-F(i-2Q,u_k)\theta(3Q-i)\theta(i-2Q) -F(i-3Q,u_k)\theta(4Q-i)\theta(i-3Q)\,.
\eeqa
Here, the index $i$ takes the values $1,\dots,4Q$. The function $\theta$ is the unitstep function. The coefficients $B_1, B_2$ and $F$ admit the usual perturbative expansion
\beqa \label{B1}
&&B_1(j,u_k) = B_{1,0}(j,u_k)+g^2\,B_{1,2}(j,u_k)+g^4\,B_{1,4}(j,u_k)+\Op(g^6)\,,\\ \label{B2}
&&B_2(j,u_k) = B_{2,0}(j,u_k)+g^2\,B_{2,2}(j,u_k)+g^4\,B_{2,4}(j,u_k)+\Op(g^6)\,,\\ \label{F}
&&F(j,u_k) = F_{0}(j,u_k)+g^2\,F_{2}(j,u_k)+g^4\,F_{4}(j,u_k)+\Op(g^6)\,.
\eeqa
In the above formulas, in view of \eqref{smatrixstructure}, the index $j$ takes the values $0,\dots,Q-1$. The explicit form of these coefficients can be found in the Appendix \ref{sec:eigenS}. The decomposition \eqref{smatrixstructure} enables a direct evaluation of the supertrace \eqref{matrixstr}
\beqa \nn
&&\textrm{str} \bigg(\prod^M_{k=1} \hat{S}^{\textrm{\tiny matrix}}_{\mathfrak{su}(2|2),Q}(q,u_k)\bigg)=\textrm{str} \bigg(\prod^M_{k=1} \hat{S}^{\textrm{\tiny diag}}_{\mathfrak{su}(2|2),Q}(q,u_k)\bigg)=\sum^{Q-1}_{j=0}
\bigg[\prod^M_{k=1}\bigg(B_{1,0}(j,u_k)\\ \nn
&&+g^2\,B_{1,2}(j,u_k)+g^4\,B_{1,4}(j,u_k)+\Op(g^6)\bigg)+\prod^M_{k=1}\bigg(B_{2,0}(j,u_k)+g^2\,B_{2,2}(j,u_k)\\ \nn
&&+g^4\,B_{2,4}(j,u_k)+\Op(g^6)\bigg)-2\prod^M_{k=1}\bigg(F_{0}(j,u_k)+g^2\,F_{2}(j,u_k)+g^4\,F_{4}(j,u_k)\\
&&+\Op(g^6)\bigg) \bigg] \,.
\eeqa
Upon expanding this to the order $\Op(g^4)$ and simultaneously inserting the one-loop Bethe roots $u^0_k$, one finds
\beq
\bigg(\textrm{str} \bigg(\prod^M_{k=1} \hat{S}^{\textrm{\tiny matrix}}_{\mathfrak{su}(2|2),Q}(q,u_k)\bigg)\bigg)_{u=u_0}=0+g^2\,S_{\boxplus}^{(2)}(M,q,Q)+g^4\,S_{\boxplus}^{(4)}(M,q,Q)+\Op(g^6)\,.
\eeq
The absence of the $\Op(g^0)$ contribution is non-trivial and is a result of certain functional relations between the coefficients in \eqref{B1}-\eqref{F}. Please note that the higher loop corrections to Bethe roots affect only $S_{\boxplus}^{(2)}(M,q,Q)$ and have already been taken into account in \eqref{Y82}.

\subsection{Calculating $Y^{(8,2)}_Q$} \label{sec:Y82}
As discussed in the beginning of section \ref{sec:wrapping}, the second contribution to the five-loop integrand, $Y^{(8,2)}_Q$, originates from the two-loop corrections to the magnon rapidities. The asymptotic all-loop Baxter equation for twist-two operators and its perturbative solution, $P_M(u)=\Baxter(u)+g^2 \, \Baxtersub(u)+\dots$, have been studied in \cite{Kotikov:2008pv}. In particular, the two-loop solution $\Baxtersub(u)$ has been derived
\begin{eqnarray}\label{Q1 full}
\Baxtersub(u)&=&4\Big(S_2(M)+4\,S_1(M)^2-2\,S_1(M)S_1(2M)\Big)\,{}_3 F_2\left(\left.
    \begin{array}{c}
        -M, \ M+1,\ \frac{1}{2}+iu \\
        1,\  1
    \end{array}
    \right| 1\right) \nonumber\\
&&+ \, 4\,S_1(M) \, \frac{\p}{\p\delta} \, {}_3 F_2 \left. \left(\left.
    \begin{array}{c}
        -M, \ M+1+2\delta,\ \frac{1}{2}+iu \\
        1+\delta,\  1
    \end{array}
    \right| 1\right) \right|_{\delta=0} \nonumber \\
&&- \,  \phantom{\gamma^{(0)}} \, \frac{\p^2}{\p\delta^2} \, {}_3 F_2  \left. \left(\left.
    \begin{array}{c}
        -M, \ M+1,\ \frac{1}{2}+iu \\
        1+\delta,\  1-\delta
    \end{array}
    \right| 1 \right) \right|_{\delta=0}   \nonumber\\
&&- \,  \phantom{\gamma^{(0)}} \frac{\p^2}{\p\delta^2}\, {}_3 F_2  \left.\left(\left.
    \begin{array}{c}
        -M, \ M+1,\ \frac{1}{2}+iu+\delta \\
        1+\delta,\  1+\delta
    \end{array}
    \right| 1\right) \right|_{\delta=0}  \,.
\end{eqnarray}
Knowing $\Baxtersub(u)$, it is straightforward to determine $Y^{(8,2)}(M,q,Q)$
\begin{eqnarray}\label{Y82}
\frac{\intd^{(8,2)}(M,q,Q)}{\intd^{(8,0)}(M,q,Q)}=2\frac{ \DBaxtersub (-\frac{i}{2}) - (-1)^M \DBaxtersub(\frac{i}{2})}{\DBaxter(-\frac{i}{2}) - (-1)^M \DBaxter(\frac{i}{2})} - \frac{R^{(2)}_M(q,Q)}{R^{(0)}_M(q,Q)} + 2\frac{T^{(2)}_M(q,Q)}{T^{(0)}_M(q,Q)}\,.
\end{eqnarray}
Here,
\begin{equation}
\frac{R^{(2)}_M(q,Q)}{R^{(0)}_M(q,Q)}=\frac{\Baxtersub(\frac{q - i(-1 + Q)}{2})}{\Baxter(\frac{q - i(-1 + Q)}{2})}+\frac{\Baxtersub(\frac{q + i(-1 + Q)}{2})}{\Baxter(\frac{q + i(-1 + Q)}{2})}+\frac{\Baxtersub(\frac{q - i(1 + Q)}{2})}{\Baxter(\frac{q - i(1 + Q)}{2})}+\frac{\Baxtersub(\frac{q + i(1 + Q)}{2})}{\Baxter(\frac{q + i(1 + Q)}{2})}\,,
\end{equation}
and
\begin{equation}
T^{(2)}_M(q,Q)=\sum_{j=0}^{Q-1}\Big( \frac{1}{2 j - i q - Q} -(-1)^M \frac{1}{2 (j + 1) - i q - Q} \Big) \Baxtersub\big(\frac{q - i (Q - 1)}{2}  + i j\big)\,.
\end{equation}

\subsection{Modification of the asymptotic Bethe ansatz} \label{sec:ABAmod}
An important correction that needs to be taken into consideration at the five-loop order is the modification of the quantisation condition. The form of this correction for an arbitrary number of magnons has been proposed in \cite{Bajnok:2008bm}. The rapidities of the individual magnons are influenced by the finite-size effects
\beq
u_j \to u_j +g^8 \,\delta u_j +\Op(g^{10})\,.
\eeq
This produces an additional contribution to the scaling dimension
\beq \label{ABAmodcontribution}
\Delta^{\textrm{ABA}}_w = g^{10} \sum^M_{j=1} E'(u_j) \delta u_j+\Op(g^{12})\,.
\eeq
Following the arguments presented in \cite{Bajnok:2009vm}, we expect the shifts $\delta u_j$ to be determined through
\beq \label{ABAmod}
\sum^M_{j=1}\bigg(\frac{\partial BY(u_k)}{\partial u_j}\bigg)_{u=u^0} \delta u_k+\Phi_k\big| _{u=u^0} =0\,,\qquad k=1,\dots,M \,,
\eeq
with
\beq
BY(u_k)=\left(\frac{u_k+\frac{i}{2}}{u_k-\frac{i}{2}}\right)^2\prod^M_{j \neq k} \frac{u_k-u_j+i}{u_k-u_j-i}\,.
\eeq
The one-loop Bethe roots $u^0_k$ are zeros of the function $\Big(BY(u_k)-1\Big)$. The quantities $\Phi_k$ entering \eqref{ABAmod} are given by
\beq \label{phi}
\Phi_k=\sum^\infty_{Q=1}\int^{\infty}_{-\infty} \frac{dq}{2\pi i} \left(\frac{z^-}{z^+}\right)^2\,\textrm{str} \bigg\{ S_Q(q,u_1)\ldots \partial S_Q(q,u_k)\ldots S_Q(q,u_M) \bigg\}\,.
\eeq
In practice, due to
\beq
\Phi_k = \big(\Phi_1\big)_{u_1 \leftrightarrow \,u_k}
\eeq
it is sufficient to calculate say $\Phi_1$.
Using the $G(q,Q)$ matrix introduced in the sub-section  \ref{sec:Y10}, one finds for $k=1$
\beqa  \nn
\textrm{str} \bigg\{ (\partial S_Q) S_Q\ldots S_Q \bigg\}&=&\frac{\partial S_0(q,Q,u_1)}{\partial q}\bigg(\prod^M_{k=2}S_0(q,Q,u_k)\bigg)\,\textrm{str} \bigg\{ S^{\textrm{\tiny diag}}_Q\ldots S^{\textrm{\tiny diag}}_Q \bigg\}\\ \nn
&&+\bigg(\prod^M_{k=1}S_0(q,Q,u_k)\bigg)\, \textrm{str} \bigg\{ (G^2 \,\partial S^{\textrm{\tiny matrix}}_Q \, (G^{-1})^2) S^{\textrm{\tiny diag}}_Q\ldots S^{\textrm{\tiny diag}}_Q \bigg\} \,.\\ \label{phistr}
\eeqa
Since all the matrices to the right of $G \, \partial S^{\textrm{\tiny matrix}}_Q\,G^{-1}$ are diagonal, only diagonal elements of $G \,\partial S^{\textrm{\tiny matrix}}_Q \, G^{-1}$ contribute to the supertrace. It should be noted that due to the tensor decomposition \eqref{tensorproduct} one has
\beqa \nn
G^2 \, \partial S^{\textrm{\tiny matrix}}_Q\, (G^{-1})^2&=&(G\,\partial \hat{S}_{\mathfrak{su}(2|2),Q}(q,u_k)\,G^{-1}) \otimes\hat{S}^{\textrm{\tiny diag}}_{\mathfrak{su}(2|2),Q}(q,u_k) \\
&&+\hat{S}^{\textrm{\tiny diag}}_{\mathfrak{su}(2|2),Q}(q,u_k) \otimes (G\,\partial \hat{S}_{\mathfrak{su}(2|2),Q}(q,u_k)\,G^{-1}) \,.
\eeqa
To calculate \eqref{phi} to five-loop order, it is sufficient to expand all quantities in \eqref{phistr}, with the exception of $(z^-/z^+)^2$, to the two-loop order. The diagonal part of  $G\partial \hat{S}_{\mathfrak{su}(2|2),Q}G^{-1}$ admits a decomposition similar to \eqref{smatrixstructure} with the corresponding functions
\beqa \label{DB1}
&&DB_1(j,u_k) = DB_{1,0}(j,u_k)+g^2\,DB_{1,2}(j,u_k)+\Op(g^4)\,,\\ \label{DB2}
&&DB_2(j,u_k) = DB_{2,0}(j,u_k)+g^2\,DB_{2,2}(j,u_k)+\Op(g^4)\,,\\ \label{DF}
&&DF(j,u_k) = DF_{0}(j,u_k)+g^2\,DF_{2}(j,u_k)+\Op(g^4)\,.
\eeqa
The index $j$ again takes the values $0, \ldots , Q-1$\,. We present the explicit form of the expansion coefficients in Appendix \ref{sec:diagdS}.

\subsection{The final result}
Having all ingredients, it is only a matter of computational effort to calculate \eqref{fterm} and \eqref{ABAmodcontribution}. As discussed in \cite{Bajnok:2009vm}, the rational integrals over $q$ may be performed by taking the residue at $q=i\,Q$. The remaining residues, when combined together, should \textit{cancel} upon summing over $Q=1,\ldots,\infty$. This is a very non-trivial occurrence which gives evidence that the $\mu$-terms are absent in the weak-coupling limit\footnote{We have explicitly checked the cancelation of the remaining poles for the first few values of $M$.}. There is also a contribution to the integrand that involves polygamma functions \eqref{polygammas}. They provide an additional source of infinitely many poles that need to be taken into account while calculating the integral! Bearing this in mind, we have used \texttt{Mathematica} to set up the rational integrands and took the residue at $q=i\,Q$\,, while the non-rational integrands were treated separately. Methods for performing sums over $Q$ have been developed in \cite{Bajnok:2009vm} and can be also applied to the current case. For higher values of $M$, however, we have found it more efficient to perform high precision computations and to use EZ-Face \cite{EZFace} to determine the transcendental structure. One starts with $Q=1$ and performs numerical integration over $q$ and repeats that procedure for $Q=2, 3 , \dots$. The series is fast convergent and the number of steps depends on the desired accuracy. For the first dozen of $M$, we have performed calculations with very high accuracy and subsequently  determined the transcendental structure \eqref{transcendental} using EZ-Face. This allowed us to conjecture the form of the functions multiplying zeta functions. For higher values of $M$ the non-rational part can be subtracted from the full result and it is sufficient to perform computations to the accuracy that allows to rationalise the results. In what follows we will discuss the intriguing structure of the wrapping correction.

We assume that the wrapping correction also preserves the reciprocity symmetry. This implies that a part of the wrapping correction may be found from the lower order results
\beq \label{gamma10P}
\Delta_{w}=\frac{1}{2} \Delta_w^{(8)}\dot\gamma_2
+\frac{1}{2}\dot \Delta_w^{(8)} \gamma_2
+\reciP_{10}^{w}\,,
\eeq
where $\gamma_2$ is the one-loop anomalous dimension
and $ \Delta_w^{(8)}$ is the four-loop wrapping correction found in \cite{Bajnok:2008qj}
\beqa
\gamma_2&=&\reciP_2=8\,\HS_1=4\,\HBS_1\,,\\
 \Delta_w^{(8)}&=&\reciP_8^{w}=-128\,
\HS_1^2 \Big(5\,\zf +4\,\HS_{-2}\,\zt +2\,\HS_5-2\,\HS_{-5}+4\,\HS_{3,-2}- 4\,\HS_{-2,-3}\nonumber\\
&&-4\,\HS_{4,1}+8\,\HS_{-2,-2,1}\Big)=
2\,\reciP_2^2 \Big(-5\,\zf +2\,\HBS_{2}\,\zt
+\left(\HBS_{2,1,2}-\HBS_{3,1,1}\right)\Big) \,. \label{P8}
\eeqa
The remaining part $\reciP_{10}^w$ should have the following transcendental structure
\beq \label{transcendental}
\reciP_{10}^{w}=\zs \,T_{\zs}
+{\zt}^2 \,T_{{\zt}^2}
+\zf \, T_{\zf}
+\zt \,T_{\zt}
+T_{\rm rational}\,,
\eeq
which is a plausible generalisation of the five-loop result for the Konishi operator,
see \cite{Bajnok:2009vm}. Applying the principle of maximal transcendentality we conclude
that the transcendentality of the components
$T_{\zs}$, $T_{\zt^2}$, $T_{\zf}$, $T_{\zt}$, $T_{\rm rational}$ should be equal to 2, 3, 4, 6 and 9 respectively. The lowest-transcendentality functions,
$T_{\zs}$ and $T_{\zt^2}$, may be deducted by inspecting first few values of $M$
\beqa
T_{\zs}&=& 13440\ S_1^2 \,,\\
T_{\zt^2}&=& -1536\ S_1^3\,.
\eeqa
The reconstruction of $T_{\zf}$ and $T_{\zt}$ requires approximately a dozen of values of \eqref{wrappingtotal}.
A careful analysis of their structure suggests that the five-loop wrapping correction should have the following structure
\beqa \label{gamma10split}
\reciP_{10}^{w}&=&2\,\reciP_2^2 \widetilde T
+2\,\reciP_2 \Big(2\,\reciP_4 +\frac{1}{16} \reciP^3_2\Big)
\Big(-5\,\zf +2\,\HBS_{2}\,\zt+\left(\HBS_{2,1,2}-\HBS_{3,1,1}\right)\Big)\,,\\ \label{Ttilde}
\widetilde T&=&
\zs \widetilde T_{\zs}
+{\zt}^2 \widetilde T_{{\zt}^2}
+\zf \widetilde T_{\zf}
+\zt \widetilde T_{\zt}
+\widetilde T_{\rm rational}\,,
\eeqa
with $\reciP_{4}$ (\ref{reciP4}) and $\widetilde T_i$ given by
\beqa
&&\reciP_{4}=8\big(\HBS_1\HBS_2-\HBS_{2,1}-\HBS_3\big)\,,\\
&&\widetilde T_{\zs}=105\,,\\
&&\widetilde T_{\zt^2}=-6\,\HBS_1\,,\\
&&\widetilde T_{\zf}=-40\,\HBS_2\,,\\
&&\widetilde T_{\zt}=4\big( 3\,\HBS_1 \HBS_{2,1} -2\,\HBS_{2,2}+2\,\HBS_{3,1}-\HBS_{2,1,1}-\HBS_4\big)\,.
\eeqa
The simplest ansatz for $\widetilde T_{\rm rational}$ requires 48 binomial sums. Thus, one needs to calculate \eqref{wrappingtotal} up to $M=48$, which turns out to be very elaborate. The most complicated part of the calculations is the determination of the matrix part of the integrand and the calculation of the modification of the asymptotic Bethe ansatz. For example, for $M=48$ this amounts to, respectively, 100 and 30 hours of computer time on a 2200 MHz PC.

At the end we have found the following result:
\beqa
\widetilde T_{\rm rational}&=&
2 \, \Big(
\HBS_1 \left(\HBS_{2,3,1}-\HBS_{3,1,2}\right)
-\HBS_{2,1,4}
+2\, \HBS_{2,2,3}
-5\, \HBS_{3,1,3}
+2\, \HBS_{3,2,2}\nonumber\\
&+&2\, \HBS_{3,3,1}
-\HBS_{4,1,2}
+\HBS_{5,1,1}
-2\, \HBS_{2,1,2,2}
+2\, \HBS_{2,1,3,1}
-2\, \HBS_{2,2,1,2}
-2\, \HBS_{2,2,2,1}\nonumber\\
&+&2\, \HBS_{2,3,1,1}
-2\, \HBS_{3,1,1,2}
+2\, \HBS_{3,1,2,1}
+2\, \HBS_{3,2,1,1}
-\HBS_{2,1,1,1,2}
+\HBS_{3,1,1,1,1}\Big)\,.\label{widetildeTr}
\eeqa
Putting \eqref{Ttilde}-\eqref{widetildeTr} into equation \eqref{gamma10split} and upon expressing the full wrapping contribution \eqref{gamma10P} in terms of the nested harmonic sums, one finds \eqref{finalwrapping}.

It is interesting to note, as follows from comparing \eqref{gamma10split} and \eqref{P8},  that the five-loop function $\reciP^{w}_{10}$ contains a part proportional to its four-loop counterpart $\reciP^{w}_{8}$ with the prefactor re-scaled as $\reciP^2_2 \to \reciP_2 \left(2 \,\reciP_4+\frac{1}{16}\reciP^3_2 \right)$. Thus, remarkably, the one-loop function $\reciP_2$ is (up to the $\reciP^3_2$ term) simply replaced by the two-loop contribution $\reciP_4$. The remaining part, on the other hand, is
proportional to $\reciP_2^2\sim\HBS_1^2$, similarly to the four-loop wrapping
correction. If a similar decomposition is also a feature of higher orders of perturbation theory, this might considerably simplify the calculation of the wrapping corrections for twist-two operators.

\subsection{Large $M$ asymptotic and analytical continuation}
In this section we check our result against the known constraints on the five-loop anomalous dimension of twist-two operators. Firstly, we find that the $M\to\infty$ limit can be easily calculated
with use of the \texttt{SUMMER} package~\cite{Vermaseren:1998uu}
for \texttt{FORM} ~\cite{Vermaseren:2000nd} giving
\beq
\lim_{M \to \infty} \Delta_w (M)=0\,.
\eeq
This means that, in similarity to the four-loop case, the wrapping effects do not influence the scaling function and may contribute starting from the order $\Op \left( \frac{\log M}{M} \right)$ only.

Secondly, we use the \texttt{SUMMER} package to transform the ABA contribution and the wrapping correction into the canonical basis. The harmonic sums corresponding to the highest poles at $M=-2+\omega$ enter with the following coefficients
\beqa
\gamma_{10}^{ABA}&=&1024 \, (13 \,\HS_9+15 \,\HS_{-9})+\ldots\,,\\\nonumber
\Delta_{w}&=&1024 \, (\HS_9-\HS_{-9})+\ldots\,.
\eeqa
Using the following analytical continuation
\beq
\HS_9(-2+\omega)=\frac{1}{\omega^9},\qquad
\HS_{-9}(-2+\omega)=\frac{1}{\omega^9},\qquad
\eeq
we confirm the result \eqref{dlevenp} following from the double-logarithmic constraints.

The most intriguing test, however, is the comparison with the predictions coming from the BFKL equation, \textit{cf.} \eqref{NLOpoles}. The expansion at $M=-1+\omega$ of the five-loop ABA result was determined in \eqref{killerexpansion}
\beqa
\gamma_{10}^{ABA}(-1+\omega)&=&
-\frac{2048}{\omega ^9}
+\frac{7168 \, \pi ^2}{3 \,\omega ^7}
-\frac{5120 \, \zeta (3)}{\omega ^6}
-\frac{63872 \, \pi ^4}{45 \, \omega ^5}\\ \nonumber
&&+512 \,\frac{13\, \pi ^2 \, \zeta (3)-678 \, \zeta (5)}{{3} \,\omega ^4}+\Op\left(\frac{1}{\omega^3}\right)\,.
\eeqa
For the five-loop wrapping correction \eqref{finalwrapping} we find
\beqa
\Delta_{w}(-1+\omega)&=&
+\frac{2048}{\omega ^9}
-\frac{7168 \, \pi ^2}{3 \, \omega ^7}
+\frac{5120 \, \zeta (3)}{\omega ^6}
+\frac{63872 \, \pi ^4}{45 \, \omega ^5}\\ \nonumber
&&-512\, \frac{5 \, \pi ^2 \, \zeta (3)-194 \, \zeta (5)}{{3}\,\omega ^4}+\Op\left(\frac{1}{\omega^3}\right)\,.
\eeqa
Summing up these two results gives
\beq
\gamma_{10}(-1+\omega)=-1024 \, \frac{2 \, \zeta(2) \, \zeta (3)+16 \, \zeta (5)}{\omega ^4}+\Op\left(\frac{1}{\omega^3}\right)\,,
\eeq
which coincides with the BFKL equation prediction \eqref{NLOpoles}!
The full agreement with the known constraints strongly corroborates our result!


 \subsection*{Acknowledgments}
We would like to thank to Fabian Spill and Arkady Tseytlin for discussions. We are also indebted to Lisa Freyhult, Matthias Staudacher and Stefan Zieme for comments on the manuscript. Tomasz \L ukowski was supported by Polish science funds during 2009-2011 as a research project (NN202 105136). T.\L.~would like to thank the Niels Bohr Institute and the Albert Einstein Institute for hospitality during the period when part of this work was performed. Adam Rej is supported by a STFC postdoctoral fellowship. Vitaly Velizhanin is supported by RFBR grants 07-02-00902-a,
RSGSS-3628.2008.2.

\appendix
\section{The ABA contribution at five-loop order} \label{sec:ABAresult}
Below we present an explicit expression for the five-loop ABA contribution to the anomalous dimension of twist-two operators $\gamma^{\textrm{ABA}}_{10}$ in terms of the usual harmonic sums \eqref{vhs}. We \textit{do not} use the canonical basis, see section \ref{sec:structure} for the definition, since this would expand the result even further.
\begin{table}[hp!]
\begin{eqnarray*}
&&\big(20480 S_{-5}-8192 S_{-3} S_{-2}+2048 S_5-20480 S_{-4,1}-16384 S_{-3,2}-\frac{28672 }{3}S_{-2,3}\\&&+\frac{32768}{3} S_{-3,1,1}+\frac{16384}{3}S_{-2,1,2}+\frac{16384}{3} S_{-2,2,1}\big) S_1^4+\big(20480 S_{-3}^2+4096 S_3^2+81920 S_{-6}\\&&+S_{-2} \big(30720 S_{-4}+8192 S_4\big)+30720 S_6-98304
   S_{-5,1}-12288 S_{-4,-2}-102400 S_{-4,2}\\&&-8192 S_{-3,-3}-90112 S_{-3,3}+S_3 \big(24576 S_{-3}-16384 S_{-2,1}\big)-57344 S_{-2,4}+4096 S_{4,2}\\&&+16384
   S_{5,1}+122880 S_{-4,1,1}-16384 S_{-3,-2,1}+106496 S_{-3,1,2}+106496 S_{-3,2,1}\\&&-16384 S_{-2,-3,1}-8192 S_{-2,-2,2}+S_2 \big(-8192 S_{-2}^2+49152 S_{-4}+8192
   S_4-\frac{131072 }{3}S_{-3,1}\\&&-\frac{81920 }{3}S_{-2,2}+\frac{65536}{3} S_{-2,1,1}\big)+65536 S_{-2,1,3}+65536 S_{-2,2,2}+65536 S_{-2,3,1}\\&&-98304
   S_{-3,1,1,1}-49152 S_{-2,1,1,2}-49152 S_{-2,1,2,1}-49152 S_{-2,2,1,1}\big) S_1^3+\big(\big(12288 S_{-3}\\&&+9216 S_3\big) S_{-2}^2+\big(53248 S_{-5}+24576
   S_5-61440 S_{-4,1}-40960 S_{-3,2}-20480 S_{-2,3}\\&&+32768 S_{-3,1,1}+16384 S_{-2,1,2}+16384 S_{-2,2,1}\big) S_{-2}+113664 S_{-7}+3072 S_7-163840 S_{-6,1}\\&&-172032
   S_{-5,2}-174080 S_{-4,3}-163840 S_{-3,4}+S_2^2 \big(36864 S_{-3}+12288 S_3-24576 S_{-2,1}\big)\\&& +\big(-12288 S_{-4}-36864 S_4\big) S_{-2,1}-118784
   S_{-2,5}+8192 S_{4,3}+8192 S_{5,2}-40960 S_{6,1}\\&&+253952 S_{-5,1,1}+24576 S_{-4,-2,1}+24576 S_{-4,1,-2}+266240 S_{-4,1,2}+266240 S_{-4,2,1}\\&&+16384 S_{-3,-3,1}-8192
   S_{-3,-2,2}+16384 S_{-3,1,-3}+249856 S_{-3,1,3}+8192 S_{-3,2,-2}\\&&+258048 S_{-3,2,2}+249856 S_{-3,3,1}-16384 S_{-2,-3,2}-16384 S_{-2,-2,3}+S_{-3} \big(14336
   S_{-4}\\&&+43008 S_4-49152 S_{-3,1}-24576 S_{-2,2}+32768 S_{-2,1,1}\big)+S_3 \big(52224 S_{-4}+12288 S_4\\&&-57344 S_{-3,1}-40960 S_{-2,2}+49152
   S_{-2,1,1}\big)+172032 S_{-2,1,4}+180224 S_{-2,2,3}\\&&+180224 S_{-2,3,2}+172032 S_{-2,4,1}-8192 S_{4,1,2}-8192 S_{4,2,1}-32768 S_{5,1,1}\\&&-368640 S_{-4,1,1,1}+32768
   S_{-3,-2,1,1}-344064 S_{-3,1,1,2}-344064 S_{-3,1,2,1}-344064 S_{-3,2,1,1}\\&&+32768 S_{-2,-3,1,1}+16384 S_{-2,-2,1,2}+16384 S_{-2,-2,2,1}+S_2 \big(92160
   S_{-5}+S_{-2} \big(49152 S_{-3}\\&&+24576 S_3\big)+30720 S_5-122880 S_{-4,1}-12288 S_{-3,-2}-122880 S_{-3,2}-86016 S_{-2,3}\\&&+12288 S_{4,1}+172032 S_{-3,1,1}-24576
   S_{-2,-2,1}+122880 S_{-2,1,2}+122880 S_{-2,2,1}\\&&-147456 S_{-2,1,1,1}\big)-221184 S_{-2,1,1,3}-221184 S_{-2,1,2,2}-221184 S_{-2,1,3,1}\\&&-221184 S_{-2,2,1,2}-221184
   S_{-2,2,2,1}-221184 S_{-2,3,1,1}+393216 S_{-3,1,1,1,1}\\&&+196608 S_{-2,1,1,1,2}+196608 S_{-2,1,1,2,1}+196608 S_{-2,1,2,1,1}+196608 S_{-2,2,1,1,1}\big)
   S_1^2\\&&+\big(2048 S_2^4+8192 S_{-2} S_2^3+\big(9216 S_{-2}^2+24576 S_{-4}+9216 S_4-36864 S_{-3,1}-30720 S_{-2,2}\\&&+49152 S_{-2,1,1}\big) S_2^2+\big(4096
   S_{-2}^3+\big(32768 S_{-4}+24576 S_4-49152 S_{-3,1}-24576 S_{-2,2}\\&&+32768 S_{-2,1,1}\big) S_{-2}+6144 S_3^2+53248 S_{-6}+6144 S_6-90112 S_{-5,1}-94208
   S_{-4,2}\\&&-94208 S_{-3,3}+S_3 \big(32768 S_{-3}-32768 S_{-2,1}\big)-16384 S_{-3} S_{-2,1}-77824 S_{-2,4}+8192 S_{4,2}\\&&-16384 S_{5,1}+163840 S_{-4,1,1}+16384
   S_{-3,-2,1}+16384 S_{-3,1,-2}+172032 S_{-3,1,2}\\&&+172032 S_{-3,2,1}-16384 S_{-2,-2,2}+139264 S_{-2,1,3}+147456 S_{-2,2,2}+139264 S_{-2,3,1}
\end{eqnarray*}
\end{table}
\newpage
\begin{table}[hp!]
\begin{eqnarray*}
&&-16384 S_{4,1,1}-294912
   S_{-3,1,1,1}+32768 S_{-2,-2,1,1}-245760 S_{-2,1,1,2}-245760 S_{-2,1,2,1}\\
&&-245760 S_{-2,2,1,1}+393216 S_{-2,1,1,1,1}\big) S_2+13824 S_{-4}^2+4608 S_4^2+16384
   S_{-3,1}^2\\
&&+14336 S_{-2,2}^2+57344 S_{-8}+S_{-2}^2 \big(3072 S_{-4}+12288 S_4\big)+64512 S_8-98304 S_{-7,1}\\
&&-30720 S_{-6,-2}-98304 S_{-6,2}-16384
   S_{-5,-3}-102400 S_{-5,3}-3072 S_{-4,-4}-98304 S_{-4,4}\\
&&-98304 S_{-3,5}-92160 S_{-2,6}-15360 S_{4,4}-12288 S_{5,3}+26624 S_{6,2}+36864 S_{7,1}\\
&&+163840
   S_{-6,1,1}-24576 S_{-5,-2,1}+180224 S_{-5,1,2}+180224 S_{-5,2,1}-24576 S_{-4,-3,1}\\
&&-6144 S_{-4,-2,-2}-18432 S_{-4,-2,2}+184320 S_{-4,1,3}+196608 S_{-4,2,2}+184320
   S_{-4,3,1}\\
&&-8192 S_{-3,-4,1}-4096 S_{-3,-3,-2}-28672 S_{-3,-3,2}-4096 S_{-3,-2,-3}+12288 S_{-3,-2,3}\\
&&+180224 S_{-3,1,4}+192512 S_{-3,2,3}+192512 S_{-3,3,2}+176128
   S_{-3,4,1}+8192 S_{-2,-5,1}\\
&&-22528 S_{-2,-4,2}+4096 S_{-2,-3,3}+30720 S_{-2,-2,4}+S_{-3,1} \big(36864 S_{-2,2}-16384 S_{-2,1,1}\big)\\
&&-8192 S_{-2,2}
   S_{-2,1,1}+S_{-4} \big(-14336 S_{-3,1}-10240 S_{-2,2}+36864 S_{-2,1,1}\big)\\
&&+S_4 \big(30720 S_{-4}-51200 S_{-3,1}-43008 S_{-2,2}+69632
   S_{-2,1,1}\big)+139264 S_{-2,1,5}\\
&&+S_{-2,1} \big(-4096 S_{-5}-20480 S_5+24576 S_{-4,1}+36864 S_{-3,2}+28672 S_{-2,3}-16384 S_{-3,1,1}\\
&&-8192 S_{-2,1,2}-8192
   S_{-2,2,1}\big)+145408 S_{-2,2,4}+147456 S_{-2,3,3}+143360 S_{-2,4,2}\\
&&+131072 S_{-2,5,1}-8192 S_{4,1,3}-8192 S_{4,2,2}-8192 S_{4,3,1}-16384 S_{5,1,2}-16384
   S_{5,2,1}\\
&&-294912 S_{-5,1,1,1}-319488 S_{-4,1,1,2}-319488 S_{-4,1,2,1}-319488 S_{-4,2,1,1}+49152 S_{-3,-3,1,1}\\
&&+8192 S_{-3,-2,-2,1}+16384 S_{-3,-2,1,2}+16384
   S_{-3,-2,2,1}-16384 S_{-3,1,1,-3}-311296 S_{-3,1,1,3}\\
&&-327680 S_{-3,1,2,2}-311296 S_{-3,1,3,1}-16384 S_{-3,2,-2,1}-327680 S_{-3,2,1,2}-327680 S_{-3,2,2,1}\\
&&-311296
   S_{-3,3,1,1}+73728 S_{-2,-4,1,1}+8192 S_{-2,-3,-2,1}+40960 S_{-2,-3,1,2}+40960 S_{-2,-3,2,1}\\
&&+8192 S_{-2,-2,-3,1}+4096 S_{-2,-2,-2,2}+16384 S_{-2,-2,1,3}+16384
   S_{-2,-2,2,2}+16384 S_{-2,-2,3,1}\\
&&-24576 S_{-2,1,1,-4}+S_{-3} \big(40960 S_{-5}+16384 S_5-28672 S_{-4,1}-22528 S_{-3,2}-22528 S_{-2,3}\\
&&+4096 S_{4,1}+49152
   S_{-3,1,1}-8192 S_{-2,-2,1}+36864 S_{-2,1,2}+36864 S_{-2,2,1}\\
&&-49152 S_{-2,1,1,1}\big)+S_3 \big(40960 S_{-5}+8192 S_5-53248 S_{-4,1}-51200
   S_{-3,2}-\frac{112640 S_{-2,3}}{3}\\
&&+\frac{212992}{3} S_{-3,1,1}+\frac{143360}{3} S_{-2,1,2}+\frac{143360}{3} S_{-2,2,1}-49152 S_{-2,1,1,1}\big)-221184
   S_{-2,1,1,4}\\
&&-8192 S_{-2,1,2,-3}-237568 S_{-2,1,2,3}-237568 S_{-2,1,3,2}-221184 S_{-2,1,4,1}-16384 S_{-2,2,-3,1}\\
&&-8192 S_{-2,2,-2,2}-8192 S_{-2,2,1,-3}+S_{-2}
   \big(4096 S_{-3}^2+8192 S_3^2+56320 S_{-6}+25600 S_6\\
&&-32768 S_{-5,1}-26624 S_{-4,2}-28672 S_{-3,3}+S_3 \big(20480 S_{-3}-8192 S_{-2,1}\big)-24576
   S_{-2,4}\\
&&+2048 S_{4,2}+8192 S_{5,1}+36864 S_{-4,1,1}-8192 S_{-3,-2,1}+36864 S_{-3,1,2}+36864 S_{-3,2,1}\\
&&-8192 S_{-2,-3,1}-4096 S_{-2,-2,2}+24576 S_{-2,1,3}+24576
   S_{-2,2,2}+24576 S_{-2,3,1}\\
&&-49152 S_{-3,1,1,1}-24576 S_{-2,1,1,2}-24576 S_{-2,1,2,1}-24576 S_{-2,2,1,1}\big)-237568 S_{-2,2,1,3}\\
&&-245760 S_{-2,2,2,2}-237568
   S_{-2,2,3,1}-16384 S_{-2,3,-2,1}-237568 S_{-2,3,1,2}-237568 S_{-2,3,2,1}\\
&&-221184 S_{-2,4,1,1}+24576 S_{4,1,1,2}+24576 S_{4,1,2,1}+24576 S_{4,2,1,1}+98304
   S_{5,1,1,1}
 \end{eqnarray*}
\end{table}
\begin{table}[hp!]
\begin{eqnarray*}
&&+491520 S_{-4,1,1,1,1}-98304 S_{-3,-2,1,1,1}-32768 S_{-3,1,-2,1,1}+491520 S_{-3,1,1,1,2}\\
&&+491520 S_{-3,1,1,2,1}+491520 S_{-3,1,2,1,1}+491520
   S_{-3,2,1,1,1}-98304 S_{-2,-3,1,1,1}\\
&&-49152 S_{-2,-2,1,1,2}-49152 S_{-2,-2,1,2,1}-49152 S_{-2,-2,2,1,1}-32768 S_{-2,1,-3,1,1}\\
&&-16384 S_{-2,1,-2,1,2}-16384
   S_{-2,1,-2,2,1}+327680 S_{-2,1,1,1,3}+327680 S_{-2,1,1,2,2}\\
&&+327680 S_{-2,1,1,3,1}+327680 S_{-2,1,2,1,2}+327680 S_{-2,1,2,2,1}+327680 S_{-2,1,3,1,1}\\
&&-16384
   S_{-2,2,-2,1,1}+327680 S_{-2,2,1,1,2}+327680 S_{-2,2,1,2,1}+327680 S_{-2,2,2,1,1}\\
&&+327680 S_{-2,3,1,1,1}-655360 S_{-3,1,1,1,1,1}-327680 S_{-2,1,1,1,1,2}-327680
   S_{-2,1,1,1,2,1}\\
&&-327680 S_{-2,1,1,2,1,1}-327680 S_{-2,1,2,1,1,1}-327680 S_{-2,2,1,1,1,1}\big) S_1+512 S_3^3-7168 S_{-9}\\
&&+7168 S_9-18432 S_{-8,1}-2048
   S_{-2,-7}+S_3^2 \big(3072 S_{-3}-2048 S_{-2,1}\big)+S_2^3 \big(1024 S_{-3}\\
&&+1024 S_3-2048 S_{-2,1}\big)+S_{-2} \big(3072 S_{-3} S_4-6144 S_{-2,1} S_4+S_3
   \big(3072 S_{-4}+6144 S_4\\
&&-4096 S_{-3,1}-2048 S_{-2,2}\big)\big)-8192 S_{1,-8}+8192 S_{1,8}-16384 S_{2,-7}+16384 S_{2,7}\\
&&-3072 S_{3,-6}+3072 S_{3,6}-13824
   S_{4,-5}+4608 S_{4,5}-34816 S_{5,-4}-2048 S_{5,4}-35328 S_{6,-3}\\
&&-4608 S_{6,3}+10240 S_{7,-2}+9216 S_{7,2}+16384 S_{8,1}+26624 S_{-7,1,1}-27648 S_{-6,-2,1}\\
&&-6144
   S_{-6,1,-2}+12288 S_{-6,1,2}+12288 S_{-6,2,1}-18432 S_{-5,-3,1}-2048 S_{-5,-2,-2}\\
&&-4096 S_{-5,-2,2}-18432 S_{-5,1,-3}-4096 S_{-5,2,-2}+26624 S_{-4,-4,1}+44032
   S_{-4,-3,-2}\\
&&+51200 S_{-4,-3,2}+70656 S_{-4,-2,-3}+12288 S_{-4,-2,3}+13312 S_{-4,1,-4}+17408 S_{-4,1,4}\\
&&+7168 S_{-4,2,-3}-1024 S_{-4,3,-2}+44032 S_{-4,4,1}-10240
   S_{-3,-5,1}+45056 S_{-3,-4,-2}\\
&&+51200 S_{-3,-4,2}+157696 S_{-3,-3,-3}+33792 S_{-3,-3,3}+73728 S_{-3,-2,-4}+8192 S_{-3,-2,4}\\
&&-8192 S_{-3,1,-5}+61440
   S_{-3,1,5}+14336 S_{-3,2,-4}+20480 S_{-3,2,4}-3072 S_{-3,3,-3}\\
&&+10240 S_{-3,4,-2}+45056 S_{-3,4,2}+90112 S_{-3,5,1}-13312 S_{-2,-6,1}+1024 S_{-2,-5,-2}\\
&&-4096
   S_{-2,-5,2}+68608 S_{-2,-4,-3}+12288 S_{-2,-4,3}+70656 S_{-2,-3,-4}+8192 S_{-2,-3,4}\\
&&+15360 S_{-2,-2,-5}+7168 S_{-2,-2,5}-7168 S_{-2,1,-6}+21504 S_{-2,1,6}
-10240 S_{-7,-2}\\
&&-13312 S_{-7,2}+16896 S_{-6,-3}-5632 S_{-6,3}+5120 S_{-5,-4}
+1024 S_{-5,4}+3584 S_{-4,-5}\\
&&-27136 S_{-4,5}+9216 S_{-3,-6}-23552S_{-3,6}
-4096
   S_{-2,2,-5}+28672 S_{-2,2,5}\\
&&+1024 S_{-2,3,4}+8192 S_{-2,4,-3}+11264 S_{-2,4,3}+13312 S_{-2,5,-2}+40960 S_{-2,5,2}\\
&&+35840 S_{-2,6,1}+40960 S_{1,-7,1}-11264
   S_{1,-6,-2}+8192 S_{1,-6,2}-32768 S_{1,-5,-3}\\
&&+4096 S_{1,-5,3}+18432 S_{1,-4,-4}+23552 S_{1,-4,4}-10240 S_{1,-3,-5}+71680 S_{1,-3,5}\\
&&-11264 S_{1,-2,-6}+25600
   S_{1,-2,6}+32768 S_{1,1,-7}-32768 S_{1,1,7}+8192 S_{1,2,-6}-8192 S_{1,2,6}\\
&&+4096 S_{1,3,-5}+35840 S_{1,4,-4}-6144 S_{1,4,4}+83968 S_{1,5,-3}+18432 S_{1,5,3}+17408
   S_{1,6,-2}\\
&&+22528 S_{1,6,2}-32768 S_{1,7,1}+14336 S_{2,-6,1}-20480 S_{2,-5,-2}-8192 S_{2,-5,2}\\
&&+22528 S_{2,-4,-3}+1024 S_{2,-4,3}+32768 S_{2,-3,-4}+30720
   S_{2,-3,4}-6144 S_{2,-2,-5}\\
&&+38912 S_{2,-2,5}+8192 S_{2,1,-6}-8192 S_{2,1,6}-4096 S_{2,2,-5}+16384 S_{2,2,5}-1024 S_{2,3,-4}\\
&&-5120 S_{2,3,4}+43008 S_{2,4,-3}+9216
   S_{2,4,3}+32768 S_{2,5,-2}+40960 S_{2,5,2}+6144 S_{2,6,1}\\
&&+2048 S_{3,-5,1}-3072 S_{3,-4,-2}-3072 S_{3,-4,2}+12288 S_{3,-3,-3}+1024 S_{3,-3,3}+5120
   S_{3,-2,-4}
\end{eqnarray*}
\end{table}

\begin{table}[hp!]
\begin{eqnarray*}
&&+7168 S_{3,-2,4}+4096 S_{3,1,-5}-1024 S_{3,2,-4}-5120 S_{3,2,4}+3072 S_{3,3,-3}+9216 S_{3,4,-2}\\
&&+9216 S_{3,4,2}+8192 S_{3,5,1}+39936 S_{4,-4,1}-6144
   S_{4,-3,-2}+31744 S_{4,-3,2}-6144 S_{4,-2,-3}\\
&&+15360 S_{4,-2,3}+32768 S_{4,1,-4}-6144 S_{4,1,4}+36864 S_{4,2,-3}+9216 S_{4,2,3}+8192 S_{4,3,-2}\\
&&+9216
   S_{4,3,2}-6144 S_{4,4,1}+86016 S_{5,-3,1}+8192 S_{5,-2,-2}+36864 S_{5,-2,2}+81920 S_{5,1,-3}\\
&&+18432 S_{5,1,3}+32768 S_{5,2,-2}+40960 S_{5,2,2}+18432
   S_{5,3,1}+50176 S_{6,-2,1}+20480 S_{6,1,-2}\\
&&+22528 S_{6,1,2}+22528 S_{6,2,1}-18432 S_{7,1,1}-24576 S_{-6,1,1,1}+8192 S_{-5,-2,1,1}\\
&&+28672 S_{-5,1,-2,1}+8192
   S_{-5,1,1,-2}-102400 S_{-4,-3,1,1}-88064 S_{-4,-2,-2,1}\\
&&-53248 S_{-4,-2,1,-2}-59392 S_{-4,-2,1,2}-59392 S_{-4,-2,2,1}-55296 S_{-4,1,-3,1}\\
&&-34816
   S_{-4,1,-2,-2}-43008 S_{-4,1,-2,2}-14336 S_{-4,1,1,-3}-2048 S_{-4,1,2,-2}-12288 S_{-4,2,-2,1}\\
&&-2048 S_{-4,2,1,-2}-102400 S_{-3,-4,1,1}-188416
   S_{-3,-3,-2,1}-126976 S_{-3,-3,1,-2}\\
&&-155648 S_{-3,-3,1,2}-155648 S_{-3,-3,2,1}-180224 S_{-3,-2,-3,1}-24576 S_{-3,-2,-2,-2}\\
&&-90112 S_{-3,-2,-2,2}-155648
   S_{-3,-2,1,-3}-36864 S_{-3,-2,1,3}-65536 S_{-3,-2,2,-2}\\
&&-81920 S_{-3,-2,2,2}-36864 S_{-3,-2,3,1}-61440 S_{-3,1,-4,1}-102400 S_{-3,1,-3,-2}\\
&&-122880
   S_{-3,1,-3,2}-159744 S_{-3,1,-2,-3}-30720 S_{-3,1,-2,3}-28672 S_{-3,1,1,-4}\\
&&-40960 S_{-3,1,1,4}-12288 S_{-3,1,2,-3}+2048 S_{-3,1,3,-2}-98304 S_{-3,1,4,1}-61440
   S_{-3,2,-3,1}\\
&&-40960 S_{-3,2,-2,-2}-49152 S_{-3,2,-2,2}-12288 S_{-3,2,1,-3}+4096 S_{-3,3,-2,1}+2048 S_{-3,3,1,-2}\\
&&-90112 S_{-3,4,1,1}+8192 S_{-2,-5,1,1}-83968
   S_{-2,-4,-2,1}-53248 S_{-2,-4,1,-2}-59392 S_{-2,-4,1,2}\\
&&-59392 S_{-2,-4,2,1}-169984 S_{-2,-3,-3,1}-24576 S_{-2,-3,-2,-2}-83968 S_{-2,-3,-2,2}\\
&&-151552
   S_{-2,-3,1,-3}-36864 S_{-2,-3,1,3}-65536 S_{-2,-3,2,-2}-81920 S_{-2,-3,2,2}\\
&&-36864 S_{-2,-3,3,1}-75776 S_{-2,-2,-4,1}-24576 S_{-2,-2,-3,-2}-79872
   S_{-2,-2,-3,2}\\
&&-24576 S_{-2,-2,-2,-3}-22528 S_{-2,-2,-2,3}-69632 S_{-2,-2,1,-4}-8192 S_{-2,-2,1,4}\\
&&-73728 S_{-2,-2,2,-3}-18432 S_{-2,-2,2,3}-16384
   S_{-2,-2,3,-2}-18432 S_{-2,-2,3,2}\\
&&-8192 S_{-2,-2,4,1}+12288 S_{-2,1,-5,1}-38912 S_{-2,1,-4,-2}-43008 S_{-2,1,-4,2}\\
&&-157696 S_{-2,1,-3,-3}-30720
   S_{-2,1,-3,3}-71680 S_{-2,1,-2,-4}-8192 S_{-2,1,-2,4}\\
&&+8192 S_{-2,1,1,-5}+S_{-4} \big(4608 S_{-5}+1536 S_5-9216 S_{-4,1}-9216 S_{-3,2}-9216 S_{-2,3}\\
&&+18432
   S_{-3,1,1}+18432 S_{-2,1,2}+18432 S_{-2,2,1}-36864 S_{-2,1,1,1}\big)+S_4 \big(4608 S_{-5}+1536 S_5\\
&&-9216 S_{-4,1}-9216 S_{-3,2}-9216 S_{-2,3}+18432
   S_{-3,1,1}+18432 S_{-2,1,2}+18432 S_{-2,2,1}\\
&&-36864 S_{-2,1,1,1}\big)+S_2^2 \big(3072 S_{-5}+1024 S_5-6144 S_{-4,1}-6144 S_{-3,2}+S_{-2} \big(2048 S_{-3}\\
&&+4096
   S_3-4096 S_{-2,1}\big)-6144 S_{-2,3}+12288 S_{-3,1,1}+12288 S_{-2,1,2}+12288 S_{-2,2,1}\\
&&-24576 S_{-2,1,1,1}\big)+S_{-2,2} \big(-3072 S_{-5}-1024 S_5+6144
   S_{-4,1}+6144 S_{-3,2}+6144 S_{-2,3}\\
&&-12288 S_{-3,1,1}-12288 S_{-2,1,2}-12288 S_{-2,2,1}+24576 S_{-2,1,1,1}\big)+S_{-3,1} \big(-6144 S_{-5}\\
&&-2048 S_5+12288
   S_{-4,1}+12288 S_{-3,2}+12288 S_{-2,3}-24576 S_{-3,1,1}-24576 S_{-2,1,2}\\
&&-24576 S_{-2,2,1}+49152 S_{-2,1,1,1}\big)-57344 S_{-2,1,1,5}-8192 S_{-2,1,2,-4}-14336
   S_{-2,1,2,4}\\
&&+4096 S_{-2,1,3,-3}-12288 S_{-2,1,4,-2}-43008 S_{-2,1,4,2}-90112 S_{-2,1,5,1}-20480 S_{-2,2,-4,1}
\end{eqnarray*}
\end{table}

\begin{table}[hp!]
\begin{eqnarray*}
&&-43008 S_{-2,2,-3,-2}-49152 S_{-2,2,-3,2}-79872
   S_{-2,2,-2,-3}-12288 S_{-2,2,-2,3}\\
&&-8192 S_{-2,2,1,-4}+S_{-3} \big(7680 S_{-6}+2560 S_6-12288 S_{-5,1}-12288 S_{-4,2}-12288 S_{-3,3}\\
&&-9216 S_{-2,4}+18432
   S_{-4,1,1}+18432 S_{-3,1,2}+18432 S_{-3,2,1}+12288 S_{-2,1,3}+12288 S_{-2,2,2}\\
&&+12288 S_{-2,3,1}-24576 S_{-3,1,1,1}-12288 S_{-2,1,1,2}-12288 S_{-2,1,2,1}-12288
   S_{-2,2,1,1}\big)\\
&&+S_3 \big(2560 S_{-3}^2-6144 S_{-2,1} S_{-3}+2048 S_{-2,1}^2+7680 S_{-6}+2560 S_6-12288 S_{-5,1}\\
&&-12288 S_{-4,2}-12288 S_{-3,3}-9216
   S_{-2,4}+18432 S_{-4,1,1}+18432 S_{-3,1,2}+18432 S_{-3,2,1}\\
&&+12288 S_{-2,1,3}+12288 S_{-2,2,2}+12288 S_{-2,3,1}-24576 S_{-3,1,1,1}-12288 S_{-2,1,1,2}\\
&&-12288
   S_{-2,1,2,1}-12288 S_{-2,2,1,1}\big)+S_{-2,1} \big(-15360 S_{-6}-5120 S_6+24576 S_{-5,1}\\
&&+24576 S_{-4,2}+24576 S_{-3,3}+18432 S_{-2,4}-36864 S_{-4,1,1}-36864
   S_{-3,1,2}\\
&&-36864 S_{-3,2,1}-24576 S_{-2,1,3}-24576 S_{-2,2,2}-24576 S_{-2,3,1}+49152 S_{-3,1,1,1}\\
&&+24576 S_{-2,1,1,2}+24576 S_{-2,1,2,1}+24576
   S_{-2,2,1,1}\big)
-14336 S_{-2,2,1,4}\\
&&-51200 S_{-2,2,4,1}+2048 S_{-2,3,-3,1}
-2048 S_{-2,3,-2,-2}-2048 S_{-2,3,-2,2}+4096 S_{-2,3,1,-3}\\
&&-4096 S_{-2,4,-2,1}-12288
   S_{-2,4,1,-2}-38912 S_{-2,4,1,2}-38912 S_{-2,4,2,1}\\
&&-81920 S_{-2,5,1,1}-16384 S_{1,-6,1,1}+40960 S_{1,-5,-2,1}+24576 S_{1,-5,1,-2}-83968 S_{1,-4,-3,1}\\
&&-51200
   S_{1,-4,-2,-2}-59392 S_{1,-4,-2,2}-28672 S_{1,-4,1,-3}+2048 S_{1,-4,1,3}-4096 S_{1,-4,2,-2}\\
&&+2048 S_{1,-4,3,1}-96256 S_{1,-3,-4,1}-129024 S_{1,-3,-3,-2}-155648
   S_{1,-3,-3,2}\\
&&-165888 S_{1,-3,-2,-3}-36864 S_{1,-3,-2,3}-51200 S_{1,-3,1,-4}-59392 S_{1,-3,1,4}\\
&&-40960 S_{1,-3,2,-3}+8192 S_{1,-3,3,-2}-96256 S_{1,-3,4,1}+8192
   S_{1,-2,-5,1}-51200 S_{1,-2,-4,-2}\\
&&-59392 S_{1,-2,-4,2}-157696 S_{1,-2,-3,-3}-36864 S_{1,-2,-3,3}-75776 S_{1,-2,-2,-4}\\
&&-8192 S_{1,-2,-2,4}+8192 S_{1,-2,1,-5}-65536
   S_{1,-2,1,5}-20480 S_{1,-2,2,-4}-26624 S_{1,-2,2,4}\\
&&+2048 S_{1,-2,3,-3}-8192 S_{1,-2,4,-2}-47104 S_{1,-2,4,2}-90112 S_{1,-2,5,1}-28672 S_{1,1,-6,1}\\
&&+40960
   S_{1,1,-5,-2}+16384 S_{1,1,-5,2}-45056 S_{1,1,-4,-3}-2048 S_{1,1,-4,3}-65536 S_{1,1,-3,-4}\\
&&-61440 S_{1,1,-3,4}+12288 S_{1,1,-2,-5}-77824 S_{1,1,-2,5}-16384
   S_{1,1,1,-6}+16384 S_{1,1,1,6}\\
&&+8192 S_{1,1,2,-5}-32768 S_{1,1,2,5}+2048 S_{1,1,3,-4}+10240 S_{1,1,3,4}-86016 S_{1,1,4,-3}\\
&&-18432 S_{1,1,4,3}-65536
   S_{1,1,5,-2}-81920 S_{1,1,5,2}-12288 S_{1,1,6,1}+16384 S_{1,2,-5,1}\\
&&-2048 S_{1,2,-4,-2}+4096 S_{1,2,-4,2}-110592 S_{1,2,-3,-3}+4096 S_{1,2,-3,3}-34816
   S_{1,2,-2,-4}\\
&&-28672 S_{1,2,-2,4}+8192 S_{1,2,1,-5}-32768 S_{1,2,1,5}-36864 S_{1,2,4,-2}-40960 S_{1,2,4,2}\\
&&-65536 S_{1,2,5,1}-2048 S_{1,3,-4,1}+10240
   S_{1,3,-3,-2}+4096 S_{1,3,-3,2}-14336 S_{1,3,-2,-3}\\
&&-4096 S_{1,3,-2,3}+2048 S_{1,3,1,-4}+10240 S_{1,3,1,4}-14336 S_{1,3,4,1}-75776 S_{1,4,-3,1}\\
&&+8192
   S_{1,4,-2,-2}-30720 S_{1,4,-2,2}-77824 S_{1,4,1,-3}-18432 S_{1,4,1,3}-32768 S_{1,4,2,-2}\\
&&-40960 S_{1,4,2,2}-18432 S_{1,4,3,1}-102400 S_{1,5,-2,1}-65536
   S_{1,5,1,-2}-81920 S_{1,5,1,2}\\
&&-81920 S_{1,5,2,1}-45056 S_{1,6,1,1}+16384 S_{2,-5,1,1}-38912 S_{2,-4,-2,1}-6144 S_{2,-4,1,-2}\\
&&-4096 S_{2,-4,1,2}-4096
   S_{2,-4,2,1}-139264 S_{2,-3,-3,1}-69632 S_{2,-3,-2,-2}-81920 S_{2,-3,-2,2}
  \end{eqnarray*}
\end{table}

\begin{table}[hp!]
\begin{eqnarray*}
&&-73728 S_{2,-3,1,-3}+4096 S_{2,-3,1,3}-16384 S_{2,-3,2,-2}+4096 S_{2,-3,3,1}-55296
   S_{2,-2,-4,1}\\
&&-69632 S_{2,-2,-3,-2}-81920 S_{2,-2,-3,2}-86016 S_{2,-2,-2,-3}-18432 S_{2,-2,-2,3}\\
&&-30720 S_{2,-2,1,-4}-32768 S_{2,-2,1,4}-28672 S_{2,-2,2,-3}+6144
   S_{2,-2,3,-2}-49152 S_{2,-2,4,1}\\
&&+16384 S_{2,1,-5,1}-2048 S_{2,1,-4,-2}+4096 S_{2,1,-4,2}-110592 S_{2,1,-3,-3}+4096 S_{2,1,-3,3}\\
&&-34816 S_{2,1,-2,-4}-28672
   S_{2,1,-2,4}+8192 S_{2,1,1,-5}-32768 S_{2,1,1,5}-36864 S_{2,1,4,-2}\\
&&-40960 S_{2,1,4,2}-65536 S_{2,1,5,1}-16384 S_{2,2,-3,-2}-8192 S_{2,2,-3,2}-65536
   S_{2,2,-2,-3}\\
&&-49152 S_{2,2,4,1}+8192 S_{2,3,-3,1}+10240 S_{2,3,-2,-2}+8192 S_{2,3,-2,2}-49152 S_{2,4,-2,1}\\
&&-36864 S_{2,4,1,-2}-40960 S_{2,4,1,2}-40960
   S_{2,4,2,1}-81920 S_{2,5,1,1}+6144 S_{3,-4,1,1}\\
&&-22528 S_{3,-3,-2,1}-2048 S_{3,-3,1,-2}-4096 S_{3,-3,1,2}-4096 S_{3,-3,2,1}-26624 S_{3,-2,-3,1}\\
&&-18432
   S_{3,-2,-2,-2}-18432 S_{3,-2,-2,2}-10240 S_{3,-2,1,-3}+2048 S_{3,-2,1,3}-2048 S_{3,-2,2,-2}\\
&&+2048 S_{3,-2,3,1}-2048 S_{3,1,-4,1}+10240 S_{3,1,-3,-2}+4096
   S_{3,1,-3,2}-14336 S_{3,1,-2,-3}\\
&&-4096 S_{3,1,-2,3}+2048 S_{3,1,1,-4}+10240 S_{3,1,1,4}-14336 S_{3,1,4,1}+8192 S_{3,2,-3,1}\\
&&+10240 S_{3,2,-2,-2}+8192
   S_{3,2,-2,2}-6144 S_{3,3,-2,1}-18432 S_{3,4,1,1}-63488 S_{4,-3,1,1}\\
&&+8192 S_{4,-2,-2,1}+4096 S_{4,-2,1,-2}-38912 S_{4,-2,1,2}-38912 S_{4,-2,2,1}-65536
   S_{4,1,-3,1}\\
&&+8192 S_{4,1,-2,-2}-24576 S_{4,1,-2,2}-73728 S_{4,1,1,-3}-18432 S_{4,1,1,3}-32768 S_{4,1,2,-2}\\
&&-40960 S_{4,1,2,2}-18432 S_{4,1,3,1}-40960
   S_{4,2,-2,1}-32768 S_{4,2,1,-2}-40960 S_{4,2,1,2}\\
&&-40960 S_{4,2,2,1}-18432 S_{4,3,1,1}-73728 S_{5,-2,1,1}-98304 S_{5,1,-2,1}-65536 S_{5,1,1,-2}\\
&&-81920
   S_{5,1,1,2}-81920 S_{5,1,2,1}-81920 S_{5,2,1,1}-45056 S_{6,1,1,1}+118784 S_{-4,-2,1,1,1}\\
&&+86016 S_{-4,1,-2,1,1}+24576 S_{-4,1,1,-2,1}+4096 S_{-4,1,1,1,-2}+311296
   S_{-3,-3,1,1,1}\\
&&+180224 S_{-3,-2,-2,1,1}+180224 S_{-3,-2,1,-2,1}+131072 S_{-3,-2,1,1,-2}+163840 S_{-3,-2,1,1,2}\\
&&+163840 S_{-3,-2,1,2,1}+163840
   S_{-3,-2,2,1,1}+245760 S_{-3,1,-3,1,1}+196608 S_{-3,1,-2,-2,1}\\
&&+122880 S_{-3,1,-2,1,-2}+147456 S_{-3,1,-2,1,2}+147456 S_{-3,1,-2,2,1}+122880 S_{-3,1,1,-3,1}\\
&&+81920
   S_{-3,1,1,-2,-2}+98304 S_{-3,1,1,-2,2}+24576 S_{-3,1,1,1,-3}+24576 S_{-3,1,2,-2,1}\\
&&+98304 S_{-3,2,-2,1,1}+24576 S_{-3,2,1,-2,1}+118784 S_{-2,-4,1,1,1}+167936S_{-2,-3,-2,1,1}\\
&&+172032 S_{-2,-3,1,-2,1}
+131072 S_{-2,-3,1,1,-2}
+163840 S_{-2,-3,1,1,2}
+163840 S_{-2,-3,1,2,1}\\
&&+163840 S_{-2,-3,2,1,1}
+159744
   S_{-2,-2,-3,1,1}+24576 S_{-2,-2,-2,-2,1}\\
&&+24576 S_{-2,-2,-2,1,-2}+77824 S_{-2,-2,-2,1,2}+77824 S_{-2,-2,-2,2,1}+163840 S_{-2,-2,1,-3,1}\\
&&+24576
   S_{-2,-2,1,-2,-2}+81920 S_{-2,-2,1,-2,2}+147456 S_{-2,-2,1,1,-3}+36864 S_{-2,-2,1,1,3}\\
&&+65536 S_{-2,-2,1,2,-2}+81920 S_{-2,-2,1,2,2}+36864 S_{-2,-2,1,3,1}+81920
   S_{-2,-2,2,-2,1}\\
&&+65536 S_{-2,-2,2,1,-2}+81920 S_{-2,-2,2,1,2}+81920 S_{-2,-2,2,2,1}+36864 S_{-2,-2,3,1,1}\\
&&+86016 S_{-2,1,-4,1,1}+192512 S_{-2,1,-3,-2,1}+122880
   S_{-2,1,-3,1,-2}\\
&&+147456 S_{-2,1,-3,1,2}+147456 S_{-2,1,-3,2,1}+176128 S_{-2,1,-2,-3,1}+24576 S_{-2,1,-2,-2,-2}\\
&&+86016 S_{-2,1,-2,-2,2}+155648
   S_{-2,1,-2,1,-3}+36864 S_{-2,1,-2,1,3}+65536 S_{-2,1,-2,2,-2}
     \end{eqnarray*}
\end{table}

\begin{table}[hp!]
\begin{eqnarray*}
&&+81920 S_{-2,1,-2,2,2}+36864 S_{-2,1,-2,3,1}+40960 S_{-2,1,1,-4,1}+86016 S_{-2,1,1,-3,-2}\\
&&+98304
   S_{-2,1,1,-3,2}+159744 S_{-2,1,1,-2,-3}+24576 S_{-2,1,1,-2,3}+16384 S_{-2,1,1,1,-4}\\
&&+28672 S_{-2,1,1,1,4}+102400 S_{-2,1,1,4,1}+32768 S_{-2,1,2,-3,1}+28672
   S_{-2,1,2,-2,-2}\\
&&+32768 S_{-2,1,2,-2,2}-8192 S_{-2,1,3,-2,1}+86016 S_{-2,1,4,1,1}+98304 S_{-2,2,-3,1,1}\\
&&+102400 S_{-2,2,-2,-2,1}+57344 S_{-2,2,-2,1,-2}+65536
   S_{-2,2,-2,1,2}+65536 S_{-2,2,-2,2,1}\\
&&+32768 S_{-2,2,1,-3,1}+28672 S_{-2,2,1,-2,-2}+32768 S_{-2,2,1,-2,2}+4096 S_{-2,3,-2,1,1}\\
&&-8192 S_{-2,3,1,-2,1}+77824
   S_{-2,4,1,1,1}+118784 S_{1,-4,-2,1,1}+49152 S_{1,-4,1,-2,1}\\
&&+8192 S_{1,-4,1,1,-2}+311296 S_{1,-3,-3,1,1}+192512 S_{1,-3,-2,-2,1}+139264 S_{1,-3,-2,1,-2}\\
&&+163840
   S_{1,-3,-2,1,2}+163840 S_{1,-3,-2,2,1}+221184 S_{1,-3,1,-3,1}+118784 S_{1,-3,1,-2,-2}\\
&&+147456 S_{1,-3,1,-2,2}+81920 S_{1,-3,1,1,-3}+8192 S_{1,-3,1,2,-2}+73728
   S_{1,-3,2,-2,1}\\
&&+8192 S_{1,-3,2,1,-2}+118784 S_{1,-2,-4,1,1}+184320 S_{1,-2,-3,-2,1}+131072 S_{1,-2,-3,1,-2}\\
&&+163840 S_{1,-2,-3,1,2}+163840 S_{1,-2,-3,2,1}+184320
   S_{1,-2,-2,-3,1}+24576 S_{1,-2,-2,-2,-2}\\
&&+94208 S_{1,-2,-2,-2,2}+155648 S_{1,-2,-2,1,-3}+36864 S_{1,-2,-2,1,3}+65536 S_{1,-2,-2,2,-2}\\
&&+81920 S_{1,-2,-2,2,2}+36864
   S_{1,-2,-2,3,1}+81920 S_{1,-2,1,-4,1}+118784 S_{1,-2,1,-3,-2}\\
&&+147456 S_{1,-2,1,-3,2}+159744 S_{1,-2,1,-2,-3}+36864 S_{1,-2,1,-2,3}+40960 S_{1,-2,1,1,-4}\\
&&+53248
   S_{1,-2,1,1,4}+24576 S_{1,-2,1,2,-3}-4096 S_{1,-2,1,3,-2}+94208 S_{1,-2,1,4,1}\\
&&+90112 S_{1,-2,2,-3,1}+53248 S_{1,-2,2,-2,-2}+65536 S_{1,-2,2,-2,2}+24576
   S_{1,-2,2,1,-3}\\
&&-4096 S_{1,-2,3,1,-2}+94208 S_{1,-2,4,1,1}-32768 S_{1,1,-5,1,1}+77824 S_{1,1,-4,-2,1}\\
&&+12288 S_{1,1,-4,1,-2}+8192 S_{1,1,-4,1,2}+8192
   S_{1,1,-4,2,1}+278528 S_{1,1,-3,-3,1}\\
&&+139264 S_{1,1,-3,-2,-2}+163840 S_{1,1,-3,-2,2}+147456 S_{1,1,-3,1,-3}-8192 S_{1,1,-3,1,3}\\
&&+32768 S_{1,1,-3,2,-2}-8192
   S_{1,1,-3,3,1}+110592 S_{1,1,-2,-4,1}+139264 S_{1,1,-2,-3,-2}\\
&&+163840 S_{1,1,-2,-3,2}+172032 S_{1,1,-2,-2,-3}+36864 S_{1,1,-2,-2,3}+61440 S_{1,1,-2,1,-4}\\
&&+65536
   S_{1,1,-2,1,4}+57344 S_{1,1,-2,2,-3}-12288 S_{1,1,-2,3,-2}+98304 S_{1,1,-2,4,1}\\
&&-32768 S_{1,1,1,-5,1}+4096 S_{1,1,1,-4,-2}-8192 S_{1,1,1,-4,2}+221184
   S_{1,1,1,-3,-3}\\
&&-8192 S_{1,1,1,-3,3}+69632 S_{1,1,1,-2,-4}+57344 S_{1,1,1,-2,4}-16384 S_{1,1,1,1,-5}+65536 S_{1,1,1,1,5}\\
&&+73728 S_{1,1,1,4,-2}+81920
   S_{1,1,1,4,2}+131072 S_{1,1,1,5,1}+32768 S_{1,1,2,-3,-2}\\
&&+16384 S_{1,1,2,-3,2}+131072 S_{1,1,2,-2,-3}+98304 S_{1,1,2,4,1}-16384 S_{1,1,3,-3,1}\\
&&-20480
   S_{1,1,3,-2,-2}-16384 S_{1,1,3,-2,2}+98304 S_{1,1,4,-2,1}+73728 S_{1,1,4,1,-2}\\
&&+81920 S_{1,1,4,1,2}+81920 S_{1,1,4,2,1}+163840 S_{1,1,5,1,1}-8192
   S_{1,2,-4,1,1}+163840 S_{1,2,-3,-2,1}\\
&&+57344 S_{1,2,-3,1,-2}+16384 S_{1,2,-3,1,2}+16384 S_{1,2,-3,2,1}+147456 S_{1,2,-2,-3,1}\\
&&+73728 S_{1,2,-2,-2,-2}+81920
   S_{1,2,-2,-2,2}+90112 S_{1,2,-2,1,-3}-8192 S_{1,2,-2,1,3}\\
&&+24576 S_{1,2,-2,2,-2}-8192 S_{1,2,-2,3,1}+32768 S_{1,2,1,-3,-2}+16384 S_{1,2,1,-3,2}\\
&&+131072
   S_{1,2,1,-2,-3}+98304 S_{1,2,1,4,1}+81920 S_{1,2,4,1,1}-8192 S_{1,3,-3,1,1}+28672 S_{1,3,-2,-2,1}\\
&&+8192 S_{1,3,-2,1,2}+8192 S_{1,3,-2,2,1}-16384
   S_{1,3,1,-3,1}-20480 S_{1,3,1,-2,-2}-16384 S_{1,3,1,-2,2}
    \end{eqnarray*}
\end{table}

\begin{table}[hp!]
\begin{eqnarray*}
&&+61440 S_{1,4,-2,1,1}+90112 S_{1,4,1,-2,1}+65536 S_{1,4,1,1,-2}+81920 S_{1,4,1,1,2}+81920
   S_{1,4,1,2,1}\\
&&+81920 S_{1,4,2,1,1}+163840 S_{1,5,1,1,1}+8192 S_{2,-4,1,1,1}+163840 S_{2,-3,-2,1,1}\\
&&+114688 S_{2,-3,1,-2,1}+32768 S_{2,-3,1,1,-2}+163840
   S_{2,-2,-3,1,1}+98304 S_{2,-2,-2,-2,1}\\
&&+73728 S_{2,-2,-2,1,-2}+81920 S_{2,-2,-2,1,2}+81920 S_{2,-2,-2,2,1}+131072 S_{2,-2,1,-3,1}\\
&&+65536 S_{2,-2,1,-2,-2}+81920
   S_{2,-2,1,-2,2}+57344 S_{2,-2,1,1,-3}+8192 S_{2,-2,1,2,-2}\\
&&+49152 S_{2,-2,2,-2,1}+8192 S_{2,-2,2,1,-2}-8192 S_{2,1,-4,1,1}+163840 S_{2,1,-3,-2,1}\\
&&+57344
   S_{2,1,-3,1,-2}+16384 S_{2,1,-3,1,2}+16384 S_{2,1,-3,2,1}+147456 S_{2,1,-2,-3,1}\\
&&+73728 S_{2,1,-2,-2,-2}+81920 S_{2,1,-2,-2,2}+90112 S_{2,1,-2,1,-3}-8192
   S_{2,1,-2,1,3}\\
&&+24576 S_{2,1,-2,2,-2}-8192 S_{2,1,-2,3,1}+32768 S_{2,1,1,-3,-2}+16384 S_{2,1,1,-3,2}\\
&&+131072 S_{2,1,1,-2,-3}+98304 S_{2,1,1,4,1}+81920
   S_{2,1,4,1,1}+16384 S_{2,2,-3,1,1}\\
&&+98304 S_{2,2,-2,-2,1}+32768 S_{2,2,-2,1,-2}+16384 S_{2,2,-2,1,2}+16384 S_{2,2,-2,2,1}\\
&&-16384 S_{2,3,-2,1,1}+81920
   S_{2,4,1,1,1}+8192 S_{3,-3,1,1,1}+36864 S_{3,-2,-2,1,1}\\
&&+16384 S_{3,-2,1,-2,1}+4096 S_{3,-2,1,1,-2}-8192 S_{3,1,-3,1,1}+28672 S_{3,1,-2,-2,1}\\
&&+8192
   S_{3,1,-2,1,2}+8192 S_{3,1,-2,2,1}-16384 S_{3,1,1,-3,1}-20480 S_{3,1,1,-2,-2}-16384 S_{3,1,1,-2,2}\\
&&-16384 S_{3,2,-2,1,1}+77824 S_{4,-2,1,1,1}+49152
   S_{4,1,-2,1,1}+81920 S_{4,1,1,-2,1}+65536 S_{4,1,1,1,-2}\\
&&+81920 S_{4,1,1,1,2}+81920 S_{4,1,1,2,1}+81920 S_{4,1,2,1,1}+81920 S_{4,2,1,1,1}+163840
   S_{5,1,1,1,1}\\
&&-327680 S_{-3,-2,1,1,1,1}-294912 S_{-3,1,-2,1,1,1}-196608 S_{-3,1,1,-2,1,1}-49152 S_{-3,1,1,1,-2,1}\\
&&-327680 S_{-2,-3,1,1,1,1}-155648
   S_{-2,-2,-2,1,1,1}-163840 S_{-2,-2,1,-2,1,1}\\
&&-163840 S_{-2,-2,1,1,-2,1}-131072 S_{-2,-2,1,1,1,-2}-163840 S_{-2,-2,1,1,1,2}-163840 S_{-2,-2,1,1,2,1}\\
&&-163840
   S_{-2,-2,1,2,1,1}-163840 S_{-2,-2,2,1,1,1}-294912 S_{-2,1,-3,1,1,1}-172032 S_{-2,1,-2,-2,1,1}\\
&&-180224 S_{-2,1,-2,1,-2,1}-131072 S_{-2,1,-2,1,1,-2}-163840
   S_{-2,1,-2,1,1,2}-163840 S_{-2,1,-2,1,2,1}\\
&&-163840 S_{-2,1,-2,2,1,1}-196608 S_{-2,1,1,-3,1,1}-204800 S_{-2,1,1,-2,-2,1}-114688 S_{-2,1,1,-2,1,-2}\\
&&-131072
   S_{-2,1,1,-2,1,2}-131072 S_{-2,1,1,-2,2,1}-65536 S_{-2,1,1,1,-3,1}-57344 S_{-2,1,1,1,-2,-2}\\
&&-65536 S_{-2,1,1,1,-2,2}+S_2 \big(\big(1024 S_{-3}+4096 S_3\big)
   S_{-2}^2+\big(11264 S_{-5}+5120 S_5-8192 S_{-4,1}\\
&&-6144 S_{-3,2}-8192 S_{-2,3}+2048 S_{4,1}+12288 S_{-3,1,1}-4096 S_{-2,-2,1}+12288 S_{-2,1,2}\\
&&+12288
   S_{-2,2,1}-24576 S_{-2,1,1,1}\big) S_{-2}+8192 S_{-7}+9216 S_7-16384 S_{-6,1}-6144 S_{-5,-2}\\
&&-16384 S_{-5,2}-1024 S_{-4,-3}-17408 S_{-4,3}-15360 S_{-3,4}-18432
   S_{-2,5}-5120 S_{4,3}\\
&&+4096 S_{5,2}+6144 S_{6,1}+32768 S_{-5,1,1}-6144 S_{-4,-2,1}+36864 S_{-4,1,2}+36864 S_{-4,2,1}\\
&&-4096 S_{-3,-3,1}-2048 S_{-3,-2,-2}-4096
   S_{-3,-2,2}+36864 S_{-3,1,3}+40960 S_{-3,2,2}\\
&&+36864 S_{-3,3,1}+2048 S_{-2,-4,1}-8192 S_{-2,-3,2}+10240 S_{-2,-2,3}+S_{-2,1} \big(-4096 S_{-4}\\
&&-8192 S_4+12288
   S_{-3,1}+16384 S_{-2,2}-8192 S_{-2,1,1}\big)+S_{-3} \big(6144 S_{-4}+3072 S_4\\
&&-6144 S_{-3,1}-4096 S_{-2,2}+12288 S_{-2,1,1}\big)+S_3 \big(10240 S_{-4}+3072
   S_4-\frac{47104 S_{-3,1}}{3}
    \end{eqnarray*}
\end{table}

\begin{table}[hp!]
\begin{eqnarray*}
&&-\frac{40960 S_{-2,2}}{3}+\frac{69632}{3} S_{-2,1,1}\big)+34816 S_{-2,1,4}+36864 S_{-2,2,3}+36864 S_{-2,3,2}\\
&&+32768 S_{-2,4,1}-4096
   S_{4,1,2}-4096 S_{4,2,1}-73728 S_{-4,1,1,1}-81920 S_{-3,1,1,2}\\
&&-81920 S_{-3,1,2,1}-81920 S_{-3,2,1,1}+24576 S_{-2,-3,1,1}+4096 S_{-2,-2,-2,1}+8192
   S_{-2,-2,1,2}\\
&&+8192 S_{-2,-2,2,1}-8192 S_{-2,1,1,-3}-73728 S_{-2,1,1,3}-81920 S_{-2,1,2,2}-73728 S_{-2,1,3,1}\\
&&-8192 S_{-2,2,-2,1}-81920 S_{-2,2,1,2}-81920
   S_{-2,2,2,1}-73728 S_{-2,3,1,1}+24576 S_{4,1,1,1}\\
&&+163840 S_{-3,1,1,1,1}-49152 S_{-2,-2,1,1,1}-16384 S_{-2,1,-2,1,1}+163840 S_{-2,1,1,1,2}\\
&&+163840
   S_{-2,1,1,2,1}+163840 S_{-2,1,2,1,1}+163840 S_{-2,2,1,1,1}-327680 S_{-2,1,1,1,1,1}\big)\\
&&-65536 S_{-2,1,2,-2,1,1}-131072 S_{-2,2,-2,1,1,1}-65536
   S_{-2,2,1,-2,1,1}-327680 S_{1,-3,-2,1,1,1}\\
&&-294912 S_{1,-3,1,-2,1,1}-147456 S_{1,-3,1,1,-2,1}-16384 S_{1,-3,1,1,1,-2}-327680 S_{1,-2,-3,1,1,1}\\
&&-188416
   S_{1,-2,-2,-2,1,1}-180224 S_{1,-2,-2,1,-2,1}-131072 S_{1,-2,-2,1,1,-2}\\
&&-163840 S_{1,-2,-2,1,1,2}-163840 S_{1,-2,-2,1,2,1}-163840 S_{1,-2,-2,2,1,1}-294912
   S_{1,-2,1,-3,1,1}\\
&&-188416 S_{1,-2,1,-2,-2,1}-131072 S_{1,-2,1,-2,1,-2}-163840 S_{1,-2,1,-2,1,2}-163840 S_{1,-2,1,-2,2,1}\\
&&-180224 S_{1,-2,1,1,-3,1}-106496
   S_{1,-2,1,1,-2,-2}-131072 S_{1,-2,1,1,-2,2}-49152 S_{1,-2,1,1,1,-3}\\
&&-49152 S_{1,-2,1,2,-2,1}-131072 S_{1,-2,2,-2,1,1}-49152 S_{1,-2,2,1,-2,1}-16384
   S_{1,1,-4,1,1,1}\\
&&-327680 S_{1,1,-3,-2,1,1}-229376 S_{1,1,-3,1,-2,1}-65536 S_{1,1,-3,1,1,-2}-327680 S_{1,1,-2,-3,1,1}\\
&&-196608 S_{1,1,-2,-2,-2,1}-147456
   S_{1,1,-2,-2,1,-2}-163840 S_{1,1,-2,-2,1,2}-163840 S_{1,1,-2,-2,2,1}\\
&&-262144 S_{1,1,-2,1,-3,1}-131072 S_{1,1,-2,1,-2,-2}-163840 S_{1,1,-2,1,-2,2}-114688
   S_{1,1,-2,1,1,-3}\\
&&-16384 S_{1,1,-2,1,2,-2}-98304 S_{1,1,-2,2,-2,1}-16384 S_{1,1,-2,2,1,-2}+16384 S_{1,1,1,-4,1,1}\\
&&-327680 S_{1,1,1,-3,-2,1}-114688
   S_{1,1,1,-3,1,-2}-32768 S_{1,1,1,-3,1,2}-32768 S_{1,1,1,-3,2,1}\\
&&-294912 S_{1,1,1,-2,-3,1}-147456 S_{1,1,1,-2,-2,-2}-163840 S_{1,1,1,-2,-2,2}-180224
   S_{1,1,1,-2,1,-3}\\
&&+16384 S_{1,1,1,-2,1,3}-49152 S_{1,1,1,-2,2,-2}+16384 S_{1,1,1,-2,3,1}-65536 S_{1,1,1,1,-3,-2}\\
&&-32768 S_{1,1,1,1,-3,2}-262144
   S_{1,1,1,1,-2,-3}-196608 S_{1,1,1,1,4,1}-163840 S_{1,1,1,4,1,1}\\
&&-32768 S_{1,1,2,-3,1,1}-196608 S_{1,1,2,-2,-2,1}-65536 S_{1,1,2,-2,1,-2}-32768
   S_{1,1,2,-2,1,2}\\
&&-32768 S_{1,1,2,-2,2,1}+32768 S_{1,1,3,-2,1,1}-163840 S_{1,1,4,1,1,1}-32768 S_{1,2,-3,1,1,1}\\
&&-163840 S_{1,2,-2,-2,1,1}-131072
   S_{1,2,-2,1,-2,1}-49152 S_{1,2,-2,1,1,-2}-32768 S_{1,2,1,-3,1,1}\\
&&-196608 S_{1,2,1,-2,-2,1}-65536 S_{1,2,1,-2,1,-2}-32768 S_{1,2,1,-2,1,2}-32768
   S_{1,2,1,-2,2,1}\\
&&-16384 S_{1,3,-2,1,1,1}+32768 S_{1,3,1,-2,1,1}-163840 S_{1,4,1,1,1,1}-163840 S_{2,-2,-2,1,1,1}\\
&&-163840 S_{2,-2,1,-2,1,1}-98304
   S_{2,-2,1,1,-2,1}-16384 S_{2,-2,1,1,1,-2}-32768 S_{2,1,-3,1,1,1}\\
&&-163840 S_{2,1,-2,-2,1,1}-131072 S_{2,1,-2,1,-2,1}-49152 S_{2,1,-2,1,1,-2}-32768
   S_{2,1,1,-3,1,1}\\
&&-196608 S_{2,1,1,-2,-2,1}-65536 S_{2,1,1,-2,1,-2}-32768 S_{2,1,1,-2,1,2}-32768 S_{2,1,1,-2,2,1}\\
&&-32768 S_{2,2,-2,1,1,1}-16384
   S_{3,1,-2,1,1,1}+32768 S_{3,1,1,-2,1,1}-163840 S_{4,1,1,1,1,1}\\
&&+327680 S_{-2,-2,1,1,1,1,1}+327680 S_{-2,1,-2,1,1,1,1}+262144 S_{-2,1,1,-2,1,1,1}
 \end{eqnarray*}
\end{table}
\newpage
\begin{table}[hp!]
\begin{eqnarray*}
&&+131072
   S_{-2,1,1,1,-2,1,1}+327680 S_{1,-2,-2,1,1,1,1}+327680 S_{1,-2,1,-2,1,1,1}\\
&&+262144 S_{1,-2,1,1,-2,1,1}+98304 S_{1,-2,1,1,1,-2,1}+327680 S_{1,1,-2,-2,1,1,1}\\
&&+327680
   S_{1,1,-2,1,-2,1,1}+196608 S_{1,1,-2,1,1,-2,1}+32768 S_{1,1,-2,1,1,1,-2}+65536 S_{1,1,1,-3,1,1,1}\\
&&+327680 S_{1,1,1,-2,-2,1,1}+262144 S_{1,1,1,-2,1,-2,1}+98304
   S_{1,1,1,-2,1,1,-2}+65536 S_{1,1,1,1,-3,1,1}\\
&&+393216 S_{1,1,1,1,-2,-2,1}+131072 S_{1,1,1,1,-2,1,-2}+65536 S_{1,1,1,1,-2,1,2}+65536 S_{1,1,1,1,-2,2,1}\\
&&+65536
   S_{1,1,2,-2,1,1,1}+65536 S_{1,2,1,-2,1,1,1}+65536 S_{2,1,1,-2,1,1,1}-131072 S_{1,1,1,1,-2,1,1,1} \\
&&+512\,\bigg(4 S_{-2,1} S_{-3}-S_{-3}^2+S_3^2-4 S_{-2,1}^2+S_1^2 \Big(2 S_{-2}^2-4 S_{-4}+6 S_4+16 S_{-3,1}+12
   S_{-2,2}\\
&&-16 S_{-2,1,1}\Big)+S_1 \Big(-2 S_{-5}-4 S_{-3} S_2+4 S_{-2} S_3+4 S_2 S_3+6 S_5+8 S_{-4,1}-4
   S_{-3,-2}\\
&&+12 S_{-3,2}+8 S_{-2} S_{-2,1}+8 S_2 S_{-2,1}+8 S_{-2,3}+4 S_{4,1}-24 S_{-3,1,1}-8
   S_{-2,-2,1}-24 S_{-2,1,2}\\
&&-24 S_{-2,2,1}+48 S_{-2,1,1,1}\Big)\bigg)\,\zeta(3)\\
&&+2560\, S_{1}(S_3 - S_{-3} + 2\,S_{-2, 1})\,\zeta(5)
\end{eqnarray*}
\caption{The ABA contribution to the five-loop anomalous dimension of twist-two operators $\gamma^{\textrm{ABA}}_{10}$.} \label{ABAresult}
\end{table}

\section{Eigenvalues of $\hat{S}_{\mathfrak{su}(2|2),Q}$} \label{sec:eigenS}
\subsection{Bosonic eigenvalues}
The bosonic eigenvalues correspond to the bosonic part of the $\hat{S}_{\mathfrak{su}(2|2),Q}$ matrix. It turns out that their structure becomes concise and transparent upon parametrizing $B_1$ with
\beq
j^+ = j-\frac{1}{2}(-2+Q+i\,q)\,,
\eeq
and $B_2$ with
\beq
j^- = j-\frac{1}{2}(Q+i\,q)\,.
\eeq
Below we present their one-, two- and three-loop expansion coefficients using $j^+$ and $j^-$ instead of $j$.
\subsubsection{One-loop}
At the one-loop order one finds
\beq \nn
B_{1,0}=B^{0}_{1,0}+B^{1}_{1,0}\,j^{+}\,,\qquad B_{2,0}=B^{0}_{2,0}+B^{1}_{2,0}\,j^{-}\,,
\eeq
with the corresponding coefficients
\beq \nn
B^{0}_{1,0}=\frac{i-2 u}{q-i Q-2 u+i}\,, \qquad B^{1}_{1,0}=\frac{2 (2 i u+1)}{(q-i Q-2 u+i) (2 u+i)}\,,
\eeq
\beq \nn
B^{0}_{2,0}=\frac{i-2 u}{q-i Q-2 u+i}\,, \qquad  B^{1}_{2,0}=\frac{2 i}{q-i Q-2 u+i}\,.
\eeq
\subsubsection{Two-loop}
At the next order inverse powers of $j^+$ and $j^-$ appear
\beq
B_{1,2}=\frac{B^{-1}_{1,2}}{j^{+}}+B^{0}_{1,2}+B^{1}_{1,2}\,j^{+}\,,\qquad B_{2,2}=\frac{B^{-1}_{2,2}}{j^{-}}+B^{0}_{2,2}+B^{1}_{2,2}\,j^{-}\,.
\eeq
The corresponding coefficients are given by
\beq \nn
B^{-1}_{1,2}=\frac{4}{(q-i Q-2 u+i) (2 u+i)}\,,\qquad B^{0}_{1,2}=\frac{8 i \left(q^2-i (i-2 u)^2 q+Q^2+(i-2 u)^2\right)}{(q+i Q) (q-i Q-2
   u+i)^2 \left(4 u^2+1\right)}\,,
\eeq
\beqa \nn
B^{1}_{1,2}&=&\frac{16}{\left(q^2+Q^2\right) (q-i Q-2 u+i)^2 (2 u-i) (2 u+i)^3}\bigg(4 u q^3 +\\ \nn
&& (-4 u^2-4 i (Q-1) u+1)q^2 +2(2 Q^2+(i-2
   u)^2 (1-2 i u)) u q \\ \nn
&&+Q (Q-2 i u-1) (4 u (u-i Q)+1)\bigg)   \,,
\eeqa

\beq \nn
B^{-1}_{2,2}=-\frac{4}{(q-i Q-2 u+i) (2 u+i)}\,,\qquad B^{1}_{2,2}=\frac{16 \left(q^2+2 (-2 i u-1) u q+Q (Q-2 i
   u-1)\right)}{\left(q^2+Q^2\right) (q-i Q-2 u+i)^2 \left(4 u^2+1\right)}\,,
\eeq
\beqa \nn
B^{0}_{2,2}&=&\frac{1}{\left(q^2+Q^2\right) (q-i Q-2 u+i)^2 \left(4
   u^2+1\right)}\Big(8 i (q^3-i \left(4 u^2+Q+1\right) q^2\\ \nn
 &&+\left(Q^2-Q-4 (Q+1) u^2+4
   i u+1\right) q-i Q \left(Q^2+(i-2
   u)^2\right))\Big)\,.
\eeqa
\subsubsection{Three-loop}
At the three-loop order one derives
\beq
B_{1,4}=\frac{B^{-3}_{1,4}}{(j^+)^3}+\frac{B^{-2}_{1,4}}{(j^+)^2}+\frac{B^{-1}_{1,4}}{j^+}+B^0_{1,4}+B^1_{1,4}\,j^+\,,
\eeq
\beq
B_{2,4}=\frac{B^{-3}_{2,4}}{(j^-)^3}+\frac{B^{-2}_{2,4}}{(j^-)^2}+\frac{B^{-1}_{2,4}}{j^-}+B^0_{2,4}+B^1_{2,4}\,j^-\,,
\eeq
with the corresponding coefficients
\beq \nn
B^{-3}_{1,4}=-\frac{4}{(q-i Q-2 u+i) (2 u+i)}\,,\qquad B^{-2}_{1,4}=\frac{16 i q}{\left(q^2+Q^2\right) (q-i Q-2 u+i) (2 u+i)}\,,
\eeq
\beqa \nn
B^{-1}_{1,4}&=&\frac{32}{\left(q^2+Q^2\right) (q-i Q-2 u+i)^2 (2 u-i) (2 u+i)^3} \Big( (-8 q u^3-4 \left(q^2+Q^2+Q\right) u^2\\ \nn
&&+2 \left(q^3-i Q
   q^2+\left(Q^2-1\right) q-i Q^3\right) u+q^2+(Q-1)
   Q)\Big)\,,
\eeqa
\beqa \nn
B^0_{1,4}&=&\frac{32i}{(q-i Q) (q+i Q)^2 (q-i Q-2 u+i)^3 (4
   u^2+1)^3}\Big((12 u^2-1) q^5 \\ \nn
&&+(-56 u^3-12 i (Q-3) u^2+10
   u+i (Q-3)) q^4+2 (16 u^4+16 i (Q-3) u^3 \\ \nn
 && +4 (Q (3 Q+4)-10)
   u^2-4 i (Q-3) u
   -Q^2+1) q^3-i ((24 u^2-2) Q^3  \\ \nn
 &&  -2 i
   (2 u-i) (20 u^2-3) Q^2+2 (i-2 u)^2 (4 u (u-i)+1) Q   \\ \nn
&&  -(2 u-i)^3
   (6 u (4 u^2+2 i u+1)-i)) q^2+((12
   u^2-1) Q^4+8 i (i-2 u)^2 u Q^3  \\ \nn
&& +2 (i-2 u)^2 (4 u (u-2 i)-1)
   Q^2+(i-2 u)^4 (4 u (u+2 i)-3) Q  \\ \nn
   && +2 (i-2 u)^4 (1-i u (4 u (u-i)+5)))
   q-i (Q-2 i u-1)^2 (Q+2 i u+1)\times \\ \nn
&&(-16 u^4
   +4 (Q (3 Q+2)-2)
   u^2-(Q-1)^2)\Big)\,,
\eeqa
\beqa \nn
B^1_{1,4}&=&\frac{64}{(q^2+Q^2)^2 (q-i Q-2 u+i)^3 (2
   u-i)^3 (2 u+i)^5} \Big(4 u (4 u (3 u-i)-3) q^6 \\ \nn
   &&
   -(8 u (Q (4 (3 i u+1) u-3 i)+u (2 u
   (9 u-8 i)-11)+4 i)+1) q^5   \\ \nn
   &&+(-96 u^5+48 i (3 Q-1) u^4
   +16 (Q-1) (3
   Q+11) u^3-8 i (Q-1) (2 Q+13) u^2 \\ \nn
   &&-4 Q (3 Q+8) u+26 u+i Q-3 i) q^4+2
   (8 u (4 (-3 i u-1) u+3 i) Q^3\\ \nn
   &&+(8 u (u (2 (8 i-9 u) u+11)-4 i)-1)
   Q^2+4 i u (4 u^2+1)^2 Q
    \\ \nn
   &&+2 u (2 u-i)^3 (4 u (u-i)+3))
   q^3+2 (2 u (4 (i-3 u) u+3) Q^4 \\ \nn
   && +(2 i u+1) (2 u (2 u (18 u-7
   i)
   -15)+i) Q^3-(2 u-i)^3 (4 u (u+3 i)+3) Q^2\\ \nn
&&+2 (-2 i u-1)^3 (4
   u^3-u-i) Q
   +i (i-2 u)^4 u (2 u+i)^2 (6 u+i)) q^2\\ \nn
   && +(8 u (4
   (-3 i u-1) u+3 i) Q^5+(8 u (u (2 (8 i-9 u) u+11) -4 i)-1) Q^4\\ \nn
   &&+8 i u
   (4 u^2+1)^2 Q^3+4 u (2 u-i)^3 (4 u (u-i)+3) Q^2
    \\ \nn
   &&+(i-2 u)^4 (2
   u+i)^2 (4 u (u+3 i)-1) Q+2 (2 i u-1)^3 u (2 u-i)^5) q  \\ \nn
   &&-Q (Q-2 i
   u-1)^2 (4 u (4 u (3 u-i)-3) Q^3+i (i-2 u)^2 (4 u (3 u+i)+1) Q^2 \\ \nn
   && -(2
   u-i)^3 (2 u+i)^2 Q-i (4
   u^2+1)^3)\Big)\,,
\eeqa
\beq \nn
B^{-3}_{2,4}=\frac{4}{(q-i Q-2 u+i) (2 u+i)}\,,\qquad B^{-2}_{2,4}=-\frac{16 i q}{\left(q^2+Q^2\right) (q-i Q-2 u+i) (2 u+i)}\,,
\eeq
\beqa \nn
B^{-1}_{2,4}&=&\frac{32}{(q^2+Q^2) (q-i
   Q-2 u+i)^2 (2 u-i) (2 u+i)^3} \Big( (8 q u^3+4 (q^2+Q^2+Q) u^2\\ \nn
   &&+2 (-q^3+i Q
   q^2-Q^2 q+q+i Q^3) u-q^2-Q^2+Q)\Big)\,,
\eeqa
\beqa \nn
B^0_{2,4}&=&\frac{32i}{(q^2+Q^2)^2 (q-i Q-2 u+i)^3
   (4 u^2+1)^3} ((12 u^2-1) q^6+(-24 u^3\\ \nn
   &&-4 i (6 Q-1) u^2+2
   u
   +i (2 Q-3)) q^5+(-96 u^4+8 i (3 Q+8) u^3\\ \nn
   &&+4 Q (3 Q+1) u^2-2 i
   Q u-Q (Q+3)-2) q^4
   -i ((48 u^2-4) Q^3\\ \nn
&& +(4 u (2 (-6 i
   u-1) u+i)+6) Q^2-4 (4 u^2+1)^2 Q   \\ \nn
&&   -(2 u-i)^3 (2 u (2 u (6 u+5
   i)-5)+i)) q^3 \\ \nn
   &&+((1-12 u^2) Q^4+(4 u (2 (6 i u+1)
   u-i)-6) Q^3  \\ \nn
&&  -32 (i-2 u)^2 u^2 Q^2+2 (2 u-i)^3 (4 u (u (4 u+7 i)-3)-3 i)
   Q\\ \nn
&& +2 (i-2 u)^4 (u (4 (3-i u) u+7 i)
   -1)) q^2-i ((24
   u^2-2) Q^5\\ \nn
&& +(2 u (2 (-6 i u-1) u+i)+3) Q^4-4 (4 u^2+1)^2
   Q^3     \\ \nn
   && +(2 u-i)^3 (2 u (2 u (2 u+7 i)-7)-5 i) Q^2+(i-2 u)^4 (1-2 i u)
   (4 u^2-3) Q    \\ \nn
  && -i (2 u-i)^5 (2 u+i)^2) q-Q (Q-2 i u-1)^2
   (Q+2 i u+1) (-16 u^4\\ \nn
   &&+4 (Q (3 Q+2)-2)
   u^2    -(Q-1)^2))
\eeqa
\beqa \nn
B^1_{2,4}&=&\frac{64}{(q^2+Q^2)^2 (q-i Q-2
   u+i)^3 (4 u^2+1)^3} \Big((12 u^2-1) q^5+(-40 u^3\\ \nn
   &&-4 i (3 Q-5) u^2+6
   u
   +i (Q-3)) q^4+2 (Q^2 (12 u^2-1)\\ \nn
   &&-2 (i-2 u)^2 u (2
   u+3 i)) q^3-2 i ((12 u^2-1) Q^3  -i (2 u-i) (20
   u^2-3) Q^2\\ \nn
&& -2 (i-2 u)^2 (u (2 u+i)-1) Q-u (2 u-i)^3 (4 u (3 u+2
   i)-1)) q^2   \\ \nn
   &&+((12 u^2-1) Q^4-4 (i-2 u)^2 u (2 u+3 i)
   Q^2+(2 u-i)^3 (2 u+i) \times
   \\ \nn
   &&(4 u (u+3 i)-1) Q-2 i (i-2 u)^4 u (2 u+i)^2)
   q\\ \nn
&&-i Q (Q-2 i u-1)^2 (-16 u^4+8 i Q u^3+4 (3 Q^2+Q-2)
   u^2\\ \nn
&&+2 i Q u
   -Q^2+Q-1)\Big)\,.
\eeqa
\subsection{Fermionic eigenvalues}
Fermionic eigenvalues are structurally simpler than the bosonic ones. In particular, they are linear functions of $j$.
\subsubsection{One-loop}
At the one-loop order one finds
\beq
F_0 = F^0_0+F^1_0\,j\,,
\eeq
with
\beq \nn
F^0_0=1\,,\qquad F^1_0=-\frac{2}{i q+Q-2 i u-1}\,.
\eeq
\subsubsection{Two-loop}
The two-loop contribution to $F$ is given by
\beq
F_2=F^0_2+F^1_2\,j\,,
\eeq
with the coefficients
\beq \nn
F^0_2=-\frac{8 i}{(q-i Q) \left(4 u^2+1\right)}\,,\qquad F^1_2=\frac{16 \left(q^2+2 (-2 i u-1) u q+Q (Q-2 i
   u-1)\right)}{\left(q^2+Q^2\right) (q-i Q-2 u+i)^2 \left(4 u^2+1\right)}\,.
\eeq

\subsubsection{Three-loop}
At the three-loop order one determines
\beq
F_4=F^0_4+F^2_4\,j\,.
\eeq
The corresponding coefficients are given by
\beqa \nn
F^0_4&=&-\frac{32 i}{(q-i Q)^2 (q+i Q) \left(4
   u^2+1\right)^3}\Big(-16 u^4+8 (q+i Q) u^3\\ \nn
   && +4 \left(3 Q^2+Q+q (3 q-i)-2\right)
   u^2
   +2 (q+i Q) u-Q^2-q (q+i)+Q-1\Big)\,,
\eeqa
\beqa \nn
F^2_4&=&\frac{64}{(q^2+Q^2)^2 (q-i Q-2
   u+i)^3 (4 u^2+1)^3} \Big((12 u^2-1) q^5\\ \nn
&& +(-40 u^3-4 i (3 Q-5) u^2    +6
   u+i (Q-3)) q^4+2 (Q^2 (12 u^2-1)\\ \nn
&&-2 (i-2 u)^2 u (2
   u+3 i)) q^3-2 i ((12 u^2-1) Q^3      -i (2 u-i) (20
   u^2-3) Q^2 \\ \nn
&& -2 (i-2 u)^2 (u (2 u+i)-1) Q-u (2 u-i)^3 (4 u (3 u+2
   i)-1)) q^2   \\ \nn
&& +((12 u^2-1) Q^4-4 (i-2 u)^2 u (2 u+3 i)
   Q^2+(2 u-i)^3 (2 u+i)\times   \\ \nn
&&   (4 u (u+3 i)-1) Q -2 i (i-2 u)^4 u (2 u+i)^2)
   q-i Q (Q-2 i u-1)^2 \times \\ \nn
 && (-16 u^4+8 i Q u^3+4 (3 Q^2+Q-2)
   u^2     +2 i Q u-Q^2+Q-1)\Big)\,.
\eeqa

\section{Diagonal elements of $G \partial \hat{S}_{\mathfrak{su}(2|2),Q} G^{-1}$}\label{sec:diagdS}
In this section we use similar notation to that in Appendix \ref{sec:eigenS}.
\subsection{Bosons}
\subsubsection{One-loop}
At the one-loop order one finds
\beq
DB_{1,0}=DB^{0}_{1,0}+DB^{1}_{1,0}\,j^+\,,\qquad DB_{2,0}=DB^{0}_{2,0}+DB^{1}_{2,0}\,j^-\,,
\eeq
with the corresponding coefficients
\beq \nn
DB^0_{1,0}=\frac{(q-i (Q-2)) (2 u-i)}{(q-i Q-2 u+i)^2 (2 u+i)}\,,\qquad DB^{1}_{1,0}=-\frac{4 i u+2}{(q-i Q-2 u+i)^2 (2 u+i)}\,,
\eeq

\beq \nn
DB^{0}_{2,0}=\frac{q-i Q}{(q-i Q-2 u+i)^2}\,,\qquad DB^{1}_{2,0}=-\frac{2 i}{(q-i Q-2 u+i)^2}\,.
\eeq
\subsubsection{Two-loop}
At the next order the structure becomes more involved
\beq
DB_{1,2}=\frac{DB^{-2}_{1,2}}{(j^+)^2}+\frac{DB^{-1}_{1,2}}{j^+}+DB^0_{1,2}+DB^1_{1,2}\,j^+\,,
\eeq
\beq
DB_{2,2}=\frac{DB^{-2}_{2,2}}{(j^-)^2}+\frac{DB^{-1}_{2,2}}{j^-}+DB^0_{2,2}+DB^1_{2,2}\,j^-\,.
\eeq
The expansion coefficients read
\beq \nn
DB^{-2}_{1,2}=\frac{2 i}{(q-i Q-2 u+i) (2 u+i)}\,,\qquad DB^{-1}_{1,2}=-\frac{4}{(q-i Q-2 u+i)^2 (2 u+i)}\,,
\eeq

\beqa \nn
DB^0_{1,2}&=&-\frac{8 i}{(q-i Q) (q+i Q)^2 (q-i Q-2 u+i)^3 (2 u-i) (2 u+i)^3} \Big(4 u q^5  \\ \nn
&&-4 i (Q-2 i u-3) u q^4+2 u (4 Q^2+2 u (2 (1-6
   i u) u-7 i)-3) q^3 \\ \nn
&& +(32 i u^5    -16 (Q-5) u^4+8 i Q u^3-4 (4
   Q^2+Q-8) u^2\\ \nn
&& -2 i (Q (4 (Q-3) Q+3)+1) u     -4 Q+3) q^2+(4 u
   Q^4\\ \nn
&&+2 (u (-8 i u^3+4 u^2-6 i u-3)+i) Q^2      +4 (i-2
   u)^2 (2 u+i) Q\\ \nn
&& -(2 u-i)^3 (2 u+i)^2) q+Q (Q-2 i u-1) (16 (Q-1)
   u^4      +8 i Q u^3\\ \nn
&&-4 (Q+2) u^2-2 i Q (2 (Q-2) Q-1) u-2
   Q-1)\Big)\,,
\eeqa
\beqa \nn
DB^1_{1,2}&=&-\frac{32}{(q^2+Q^2)^2 (q-i Q-2 u+i)^3 (2 u-i) (2
   u+i)^3} (2 u q^5 \\ \nn
&& +(1-2 i (Q-1) u) q^4   +(4 Q^2+3 (i-2 u)^2 (1-2 i
   u)) u q^3\\ \nn
&& +(Q-2 i u-1) (-4 i u Q^2+(8 u^2+2) Q     -(i-2
   u)^2 u (2 u+i)) q^2\\ \nn
&& +Q (-8 i (Q-2) u^4+4 (4-3 Q) u^3+2 i Q
   u^2    +(2 Q^3-3 Q+4) u\\ \nn
   &&+i (Q-1)) q+Q^2 (Q-2 i u-1) (-2
   i u Q^2+4 u^2 Q   \\ \nn
&&+Q+(i-2 u)^2 u (2
   u+i)))\,,
\eeqa

\beq \nn
DB^{-2}_{2,2}=-\frac{2 i}{(q-i Q-2 u+i) (2 u+i)}\,,\qquad DB^{-1}_{2,2}=\frac{4}{(q-i Q-2 u+i)^2 (2 u+i)}\,,
\eeq
\beqa \nn
DB^0_{2,2}&=&-\frac{8 i }{(q^2+Q^2)^2 (q-i Q-2 u+i)^3 (4
   u^2+1)}(2 q^5-2 i (Q+\\ \nn
&&u (6 u-i)+1) q^4     +(8 i u^3-4 (4 Q+1)
   u^2+10 i u+4 Q^2-4 Q+3) q^3 \\ \nn
&&  +(-4 i Q^3+(2 i-4 u) Q^2  +(2 u-i)^3
   Q+(2 u-i)^3) q^2+Q (2 Q^3\\ \nn
&& -i (2 u-i)^3 Q+2 (-2 i u-1)^3)
   q    -i Q^2 (Q-2 i u-1) (2 Q^2+2 Q\\ \nn
&& +4 (Q+1) u^2+2 i (Q-2)
   u-1))\,,
\eeqa
\beqa \nn
DB^1_{2,2}&=&-\frac{32}{(q^2+Q^2)^2 (q-i Q-2 u+i)^3 (4
   u^2+1)} \Big(q^4+3 (-2 i u-1) u q^3  \\ \nn
&&   +(Q-2 i u-1) (2 Q+(i-2 u) u) q^2+Q (2
   u-i)\times  \\ \nn
&&  (-Q-i (Q-2) u+1) q  +Q^2 (Q-2 i u-1) (Q+u (2
   u-i))\Big)\,.
\eeqa
\subsection{Fermions}
The diagonal elements corresponding to the fermionic subspace are again simpler.
\subsubsection{One-loop}
At the one-loop order one finds
\beq
DF_0 = DF^1_0\,j\,,
\eeq
\beq \nn
DF^1_0=-\frac{2 i}{(q-i Q-2 u+i)^2}\,.
\eeq
\subsection{Two-loop}
At the next order one derives
\beq
DF_2 = DF^0_2+DF^1_2\,j\,,
\eeq
with the following coefficients
\beq \nn
DF^0_2=\frac{8 i}{(q-i Q)^2 \left(4 u^2+1\right)}\,,
\eeq
\beqa \nn
DF^1_2&=&\frac{32}{(q^2+Q^2)^2 (-q+i (Q-1)+2 u)^3 (4
   u^2+1)^2} (u q^5+(2 u^2
    \\ \nn
&& -i (Q-1) u+1) q^4   +(2 Q^2+3 (i-2
   u)^2 (1-2 i u)) u q^3\\ \nn
&& +(Q-2 i u-1) (-2 i u Q^2+(8
   u^2+2) Q     -(i-2 u)^2 u (2 u+i)) q^2\\ \nn
&&+Q (-8 i (Q-2) u^4+4
   (4-3 Q) u^3+2 i Q u^2     +(Q^3-3 Q+4) u+i (Q-1)) q\\ \nn
&&+Q^2 (Q-2
   i u-1) (-i u Q^2+4 u^2 Q      +Q+(i-2 u)^2 u (2
   u+i)))\,.
\eeqa
\section{Binomial sums} \label{app:rrsums}
We define the binomial sums $\HBS_{i_1,\ldots,i_k}$ through (see \cite{Vermaseren:1998uu})
\beq
\HBS_{i_1,\ldots,i_k}(N)=(-1)^N\sum_{j=1}^{N}(-1)^j\binom{N}{j}\binom{N+j}{j}\HS_{i_1,...,i_k}(j)\,,
\eeq
where $\HS_{i_1, \ldots ,i_k}$ is the nested harmonic sum defined in \eqref{vhs}. The advantage of
this basis is that only \textit{positive} values of the indices $i_1,\ldots,i_k$ need to be
considered. Moreover, these binomial sums appear in the real diagram calculations of the anomalous
dimensions and coefficient functions in QCD \cite{Moch:2004pa,Vogt:2004mw} and in the solution of the asymptotic
all-loop Baxter equation for twist-two operators \cite{Kotikov:2008pv}. At the five-loop order there
are 256 such sums, which is also the number of reciprocity-respecting harmonic sums defined in
\cite{Beccaria:2009vt}\footnote{Relations between the binomial and the nested harmonic sums together
with relations between the binomial and the reciprocity-respecting harmonic sums can be found
under \href{http://thd.pnpi.spb.ru/~velizh/5loop/}{\texttt{http://thd.pnpi.spb.ru/\textasciitilde velizh/5loop/}}
}. We found the basis of binomial sums much easier to handle and implement,
though a rigorous proof of the equivalence of the two basis of sums is unknown to us.


\begin{thebibliography}{99}

\bibitem{Lipatov:1993yb}
  L.~N.~Lipatov,
  {\it ``High-energy asymptotics of multicolor QCD and exactly solvable lattice models,''}
  (unpublished),
  {\tt arXiv:hep-th/9311037}.

\bibitem{Lipatov:1994xy}
  L.~N.~Lipatov,
  {\it ``Asymptotic behavior of multicolor QCD at high energies in
  connection with exactly solvable spin models,''}
  JETP Lett.\  {\bf 59} (1994) 596
  [Pisma Zh.\ Eksp.\ Teor.\ Fiz.\  {\bf 59} (1994) 571].

\bibitem{Faddeev:1994zg}
  L.~D.~Faddeev and G.~P.~Korchemsky,
  {\it ``High-energy QCD as a completely integrable model,''}
  Phys.\ Lett.\  B {\bf 342} (1995) 311,
  {\tt arXiv:hep-th/9404173}.

\bibitem{Lipatov:1976zz}
  L.~N.~Lipatov,
  {\it ``Reggeization of the vector meson and the
  vacuum singularity in nonabelian gauge theories,''}
  Sov.\ J.\ Nucl.\ Phys.\  {\bf 23} (1976) 338.

\bibitem{Kuraev:1977fs}
  E.~A.~Kuraev, L.~N.~Lipatov and V.~S.~Fadin,
  {\it ``The Pomeranchuk singularity in nonabelian gauge theories,''}
  Sov.\ Phys.\ JETP {\bf 45} (1977) 199.

\bibitem{Balitsky:1978ic}
  I.~I.~Balitsky and L.~N.~Lipatov,
  {\it ``The Pomeranchuk singularity in Quantum Chromodynamics,''}
  Sov.\ J.\ Nucl.\ Phys.\  {\bf 28} (1978) 822.

\bibitem{Kotikov:2002ab}
  A.~V.~Kotikov and L.~N.~Lipatov,
  {\it ``DGLAP and BFKL equations in the N = 4 supersymmetric
  gauge theory,''}
  Nucl.\ Phys.\  B {\bf 661} (2003) 19
  [Erratum-ibid.\  B {\bf 685} (2004) 405],
  {\tt arXiv:hep-ph/0208220}.

\bibitem{Braun:1998id}
  V.~M.~Braun, S.~E.~Derkachov and A.~N.~Manashov,
 {\it ``Integrability of three-particle evolution equations in {QCD},''}
  Phys.\ Rev.\ Lett.\  {\bf 81} (1998) 2020,
 {\tt arXiv:hep-ph/9805225}.

\bibitem{Belitsky:2005bu}
  A.~V.~Belitsky, G.~P.~Korchemsky and D.~Mueller,
  {\it ``Integrability of two-loop dilatation operator in gauge theories,''}
  Nucl.\ Phys.\  B {\bf 735} (2006) 17,
 {\tt arXiv:hep-th/0509121}.

\bibitem{Minahan:2002ve}
  J.~A.~Minahan and K.~Zarembo,
  {\it ``The Bethe ansatz for $\mathcal{N}= 4$ super Yang-Mills'',}
  JHEP {\bf 0303} (2003) 013,
  {\tt arXiv:hep-th/0212208}.

\bibitem{Beisert:2003jj}
  N.~Beisert,
 {\it  ``The complete one-loop dilatation operator of $\mathcal{N}= 4$ super Yang-Mills
  theory,''}
  Nucl.\ Phys.\  B {\bf 676} (2004) 3,
  {\tt arXiv:hep-th/0307015}.

\bibitem{Beisert:2003yb}
  N.~Beisert and M.~Staudacher,
  {\it ``The $\mathcal{N}= 4$ SYM integrable super spin chain,''}
  Nucl.\ Phys.\  B {\bf 670} (2003) 439,
  {\tt arXiv:hep-th/0307042}.

\bibitem{fussnote}
Please see \cite{Beisert:2005fw} for further references.

\bibitem{Beisert:2005fw}
N.~Beisert and M.~Staudacher, {\it ``Long-range
$\mathfrak{psu}(2,2|4)$ Bethe ans\"atze for gauge theory and
strings'',} Nucl.\ Phys.\ B {\bf 727} (2005) 1,
{\tt arXiv:hep-th/0504190}.

\bibitem{Janik:2006dc}
  R.~A.~Janik,
  {\it ``The AdS(5) x S**5 superstring worldsheet S-matrix and crossing  symmetry,''}
  Phys.\ Rev.\  D {\bf 73} (2006) 086006,
  {\tt arXiv:hep-th/0603038}.

\bibitem{Beisert:2006ez}
  N.~Beisert, B.~Eden and M.~Staudacher,
  {\it ``Transcendentality and crossing,''}
  J.\ Stat.\ Mech.\  {\bf 0701} (2007) P021,
  {\tt arXiv:hep-th/0610251}.

\bibitem{Kotikov:2007cy}
  A.~V.~Kotikov, L.~N.~Lipatov, A.~Rej, M.~Staudacher and V.~N.~Velizhanin,
  {\it ``Dressing and Wrapping,''}
  J.\ Stat.\ Mech.\  {\bf 0710} (2007) P10003,
  {\tt arXiv:0704.3586 [hep-th]}.

\bibitem{Sieg:2005kd}
  C.~Sieg and A.~Torrielli,
  {\it``Wrapping interactions and the genus expansion of the 2-point function  of
  composite operators,''}
  Nucl.\ Phys.\  B {\bf 723} (2005) 3\,,
  {\tt arXiv:hep-th/0505071}.

\bibitem{Bajnok:2008bm}
  Z.~Bajnok and R.~A.~Janik,
  {\it``Four-loop perturbative Konishi from strings and finite size effects for
  multiparticle states,''}
  Nucl.\ Phys.\  B {\bf 807} (2009) 625\,,
  {\tt arXiv:0807.0399 [hep-th]}.

\bibitem{Janik:2007wt}
  R.~A.~Janik and T.~Lukowski,
  {\it``Wrapping interactions at strong coupling -- the giant magnon,''}
  Phys.\ Rev.\  D {\bf 76} (2007) 126008\,,
  {\tt arXiv:0708.2208 [hep-th]}.

\bibitem{Fiamberti:2007rj}
  F.~Fiamberti, A.~Santambrogio, C.~Sieg and D.~Zanon,
  {\it ``Wrapping at four loops in $\mathcal{N}= 4$ SYM,''}
  {\tt arXiv:0712.3522 [hep-th]}.

\bibitem{Velizhanin:2008jd}
  V.~N.~Velizhanin,
  {\it ``The Four-Loop Konishi in $\mathcal{N}= 4$ SYM,''}
  {\tt arXiv:0808.3832 [hep-th].}

\bibitem{Bajnok:2008qj}
Z.~Bajnok, R.~A.~Janik and T.~Lukowski,
{\it ``Four loop twist two, BFKL, wrapping and strings,''}
{\tt arXiv:0811.4448 [hep-th].}

\bibitem{Bajnok:2009vm}
  Z.~Bajnok, A.~Hegedus, R.~A.~Janik and T.~Lukowski,
  {\it``Five loop Konishi from AdS/CFT,''}
  {\tt arXiv:0906.4062 [hep-th].}

\bibitem{Ambjorn:2005wa}
  J.~Ambjorn, R.~A.~Janik and C.~Kristjansen,
  {\it ``Wrapping interactions and a new source of corrections to the spin-chain  /
  string duality,''}
  Nucl.\ Phys.\  B {\bf 736} (2006) 288\,,
  {\tt arXiv:hep-th/0510171}.

\bibitem{Arutyunov:2007tc}
  G.~Arutyunov and S.~Frolov,
  {\it ``On String S-matrix, Bound States and TBA,''}
  JHEP {\bf 0712} (2007) 024\,,
  {\tt arXiv:0710.1568 [hep-th]}.


\bibitem{Arutyunov:2008zt}
  G.~Arutyunov and S.~Frolov,
  {\it``The S-matrix of String Bound States,''}
  Nucl.\ Phys.\  B {\bf 804} (2008) 90\,,
  {\tt arXiv:0803.4323 [hep-th]}.


\bibitem{Gromov:2008gj}
  N.~Gromov, V.~Kazakov and P.~Vieira,
  {\it``Finite Volume Spectrum of 2D Field Theories from Hirota Dynamics'',}
  {\tt arXiv:0812.5091 [hep-th]}.

\bibitem{Arutyunov:2009zu}
  G.~Arutyunov and S.~Frolov,
  {\it ``String hypothesis for the $AdS_5 x S^5$ mirror,''}
  JHEP {\bf 0903} (2009) 152\,,
  {\tt arXiv:0901.1417 [hep-th]}.

\bibitem{Gromov:2009tv}
  N.~Gromov, V.~Kazakov and P.~Vieira,
  {\it``Integrability for the Full Spectrum of Planar AdS/CFT,''}
  {\tt arXiv:0901.3753 [hep-th]}.

\bibitem{Bombardelli:2009ns}
  D.~Bombardelli, D.~Fioravanti and R.~Tateo,
  {\it``Thermodynamic Bethe Ansatz for planar AdS/CFT: a proposal,''}
  {\tt arXiv:0902.3930 [hep-th]}.

\bibitem{Gromov:2009bc}
  N.~Gromov, V.~Kazakov, A.~Kozak and P.~Vieira,
  {\it``Integrability for the Full Spectrum of Planar AdS/CFT II,''}
  {\tt arXiv:0902.4458 [hep-th]}.

\bibitem{Arutyunov:2009ur}
  G.~Arutyunov and S.~Frolov,
  {\it``Thermodynamic Bethe Ansatz for the $AdS_5 x S^5$ Mirror Model,''}
  JHEP {\bf 0905}, 068 (2009)\,,
  {\tt arXiv:0903.0141 [hep-th]}.

\bibitem{Gromov:2009zb}
  N.~Gromov, V.~Kazakov and P.~Vieira,
  {\it ``Exact AdS/CFT spectrum: Konishi dimension at any coupling,''}
 {\tt arXiv:0906.4240 [hep-th]}.

\bibitem{Roiban:2009aa}
  R.~Roiban and A.~A.~Tseytlin,
  {\it ``Quantum strings in $AdS_5 x S^5$: strong-coupling corrections to dimension of
  Konishi operator,''}
  {\tt arXiv:0906.4294 [hep-th]}.

\bibitem{Rej:2009dk}
  A.~Rej and F.~Spill,
  {\it ``Konishi at strong coupling from ABE,''}
  {\tt arXiv:0907.1919 [hep-th].}

\bibitem{Arutyunov:2009ax}
  G.~Arutyunov, S.~Frolov and R.~Suzuki,
  \textit{``Exploring the mirror TBA,''}
  {\tt arXiv:0911.2224 [hep-th]}.

\bibitem{Staudacher:2004tk}
  M.~Staudacher,
  {\it ``The factorized S-matrix of CFT/AdS,''}
  JHEP {\bf 0505} (2005) 054,
  {\tt arXiv:hep-th/0412188}.

\bibitem{Kotikov:2004ss}
A.~V.~Kotikov, L.~N.~Lipatov, A.~I.~Onishchenko and V.~N.~Velizhanin,
  {\it``Three-loop universal anomalous dimension of the Wilson operators in $\superN=4$ SUSY Yang-Mills model,''}
  Phys.~Lett.~B \textbf{595} (2004) 521\,,
  {\tt hep-th/0404092}.

\bibitem{Vermaseren:1998uu}
 J.~A.~M.~Vermaseren,
 {\it ``Harmonic sums, Mellin transforms and integrals,''}
 Int.\ J.\ Mod.\ Phys.\  A {\bf 14} (1999) 2037\,,
 {\tt arXiv:hep-ph/9806280}.

\bibitem{Kotikov:2003fb}
  A.~V.~Kotikov, L.~N.~Lipatov and V.~N.~Velizhanin,
  {\it `` Anomalous dimensions of Wilson operators in $\superN = 4$ SYM theory,''}
  Phys.\ Lett.\  B {\bf 557} (2003) 114\,,
  {\tt arXiv:hep-ph/0301021}.

\bibitem{Kotikov:2008pv}
  A.~V.~Kotikov, A.~Rej and S.~Zieme,
  {\it``Analytic three-loop Solutions for N=4 SYM Twist Operators,''}
  Nucl.\ Phys.\  B {\bf 813}, 460 (2009)\,,
  {\tt arXiv:0810.0691 [hep-th]}.

\bibitem{Dokshitzer:2005bf}
  Yu.~L.~Dokshitzer, G.~Marchesini and G.~P.~Salam,
  {\it ``Revisiting parton evolution and the large-x limit,''}
  Phys.\ Lett.\  B {\bf 634} (2006) 504,
  {\tt arXiv:hep-ph/0511302}.

\bibitem{Dokshitzer:2006nm}
  Yu.~L.~Dokshitzer and G.~Marchesini,
  {\it ``N = 4 SUSY Yang-Mills: Three loops made simple(r),''}
  Phys.\ Lett.\  B {\bf 646} (2007) 189\,,
  {\tt arXiv:hep-th/0612248}.

\bibitem{Basso:2006nk}
  B.~Basso and G.~P.~Korchemsky,
  {\it ``Anomalous dimensions of high-spin operators beyond
  the leading order,''}
  {\tt arXiv:hep-th/0612247}.

\bibitem{Beccaria:2009vt}
  M.~Beccaria and V.~Forini,
  {\it``Four loop reciprocity of twist two operators in N=4 SYM,''}
  JHEP {\bf 0903} (2009) 111\,,
  {\tt arXiv:0901.1256 [hep-th]}.

\bibitem{Beccaria:2007pb}
  M.~Beccaria,
  \textit{``Three loop anomalous dimensions of twist-3 gauge operators in N=4 SYM,''}
  JHEP {\bf 0709} (2007) 023\,,
  {\tt arXiv:0707.1574 [hep-th]}.

\bibitem{Beccaria:2009eq}
  M.~Beccaria, V.~Forini, T.~Lukowski and S.~Zieme,
  {\it ``Twist-three at five loops, Bethe Ansatz and wrapping,''}
  JHEP {\bf 0903} (2009) 129\,,
  {\tt arXiv:0901.4864 [hep-th]}.

\bibitem{Fadin:1998py}
  V.~S.~Fadin and L.~N.~Lipatov,
  {\it ``BFKL pomeron in the next-to-leading approximation,''}
  Phys.\ Lett.\  B {\bf 429} (1998) 127,
  {\tt arXiv:hep-ph/9802290}.

\bibitem{Kotikov:2000pm}
  A.~V.~Kotikov and L.~N.~Lipatov,
  {\it ``NLO corrections to the BFKL equation in QCD and
  in supersymmetric gauge theories,''}\\
  Nucl.\ Phys.\  B {\bf 582} (2000) 19\,,
  {\tt arXiv:hep-ph/0004008}.

\bibitem{Kotikov:2005gr}
  A.~V.~Kotikov and V.~N.~Velizhanin,
  {\it ``Analytic continuation of the Mellin moments of deep
  inelastic structure functions,''}
  {\tt arXiv:hep-ph/0501274}.

\bibitem{Gorshkov:1966ht}
  V.~G.~Gorshkov, V.~N.~Gribov, L.~N.~Lipatov and G.~V.~Frolov,
  {\it ``Doubly logarithmic asymptotic behavior in
  Quantum Electrodynamics,''}
  Sov.\ J.\ Nucl.\ Phys.\  {\bf 6} (1968) 95.

\bibitem{Gorshkov:1966hu}
  V.~G.~Gorshkov, V.~N.~Gribov, L.~N.~Lipatov and G.~V.~Frolov,
  {\it ``Backward Electron - Positron scattering at high energies,''}
  Sov.\ J.\ Nucl.\ Phys.\  {\bf 6} (1968) 262.

\bibitem{Kirschner:1982xw}
  R.~Kirschner and L.~N.~Lipatov,
  {\it ``Doubly logarithmic asymptotic of the quark
  scattering amplitude with
  nonvacuum exchange in the t-channel,''}
  Sov.\ Phys.\ JETP {\bf 56} (1982) 266.

\bibitem{Kirschner:1982qf}
  R.~Kirschner and L.~N.~Lipatov,
  {\it ``Double logarithmic asymptotics of quark scattering
  amplitudes with flavor exchange,''}
  Phys.\ Rev.\  D {\bf 26} (1982) 1202.

\bibitem{Kirschner:1983di}
  R.~Kirschner and L.~N.~Lipatov,
  {\it ``Double logarithmic asymptotics and Regge singularities
  of quark amplitudes with flavor exchange,''}
  Nucl.\ Phys.\  B {\bf 213} (1983) 122.

\bibitem{Bartels:1995iu}
  J.~Bartels, B.~I.~Ermolaev and M.~G.~Ryskin,
  {\it ``Nonsinglet contributions to the structure function $g_1$ at
  small x,''}
  Z.\ Phys.\  C {\bf 70}  (1996) 273,
  {\tt arXiv:hep-ph/9507271}.

\bibitem{Bartels:1996wc}
  J.~Bartels, B.~I.~Ermolaev and M.~G.~Ryskin,
  {\it ``Flavor singlet contribution to the structure function $g_1$ at
  small x,''}
  Z.\ Phys.\  C {\bf 72} (1996) 627,
  {\tt arXiv:hep-ph/9603204}.


\bibitem{Beccaria:2009rw}
  M.~Beccaria, A.~V.~Belitsky, A.~V.~Kotikov and S.~Zieme,
  {\it ``Analytic solution of the multiloop Baxter equation,''}
  {\tt arXiv:0908.0520 [hep-th].}

\bibitem{EZFace}
See \href{http://oldweb.cecm.sfu.ca/projects/EZFace/}{\texttt{http://oldweb.cecm.sfu.ca/projects/EZFace/}}\,.

\bibitem{Remiddi:1999ew}
  E.~Remiddi and J.~A.~M.~Vermaseren,
  \textit{``Harmonic polylogarithms,''}
  Int.\ J.\ Mod.\ Phys.\  A {\bf 15} (2000) 725\,,
  {\tt arXiv:hep-ph/9905237}.

\bibitem{Vermaseren:2000nd}
  J.~A.~M.~Vermaseren,
  \textit{``New features of FORM,''}
  {\tt arXiv:math-ph/0010025}.



\bibitem{Maitre:2005uu}
  D.~Maitre,
  \textit{``HPL, a Mathematica implementation of the harmonic polylogarithms,''}
  Comput.\ Phys.\ Commun.\  {\bf 174}, 222 (2006)\,,
  {\tt arXiv:hep-ph/0507152}.

\bibitem{Dippel}
 V.~Dippel, unpublished result.

\bibitem{Moch:2004pa}
 S.~Moch, J.~A.~M.~Vermaseren and A.~Vogt,
 {\it ``The three-loop splitting functions in QCD: The non-singlet case,''}
 Nucl.\ Phys.\  B {\bf 688} (2004) 101,
 {\tt arXiv:hep-ph/0403192}.

\bibitem{Vogt:2004mw}
  A.~Vogt, S.~Moch and J.~A.~M.~Vermaseren,
  \textit{``The three-loop splitting functions in QCD: The singlet case,''}
  Nucl.\ Phys.\  B {\bf 691} (2004) 129\,,
  {\tt arXiv:hep-ph/0404111}.

\end{thebibliography}
\end{document}